\documentclass{article}
\usepackage{amssymb,amsmath}
\usepackage{epsfig}
\usepackage{pdflscape}
\usepackage{subfigure}
\usepackage{color}
\usepackage{lscape}
\usepackage{bbm}
\usepackage{bm}
\usepackage{mathrsfs}
\usepackage{amsfonts}
\usepackage{graphics}
\usepackage{indentfirst}
\usepackage[square,numbers,sort&compress]{natbib}
\usepackage{multicol}

\newtheorem{rem}{Remark}[section]
\newcommand{\br}{\begin{rem}}
\newcommand{\er}{\end{rem}}
\newtheorem{ex}[rem]{Example}
\newcommand{\bex}{\begin{ex}}
\newcommand{\eex}{\end{ex}}
\newtheorem{Def}[rem]{Definition}
\newcommand{\bd}{\begin{Def}}
\newcommand{\ed}{\end{Def}}
\newtheorem{theorem}[rem]{Theorem}
\newcommand{\bt}{\begin{theorem}}
\newcommand{\et}{\end{theorem}}
\newtheorem{Prop}[rem]{Proposition}
\newcommand{\bp}{\begin{Prop}}
\newcommand{\ep}{\end{Prop}}
\newtheorem{lemma}[rem]{Lemma}
\newcommand{\bl}{\begin{lemma}}
\newcommand{\el}{\end{lemma}}
\newcommand{\be}{\begin{equation}}
\newcommand{\ee}{\end{equation}}
\newcommand{\bea}{\begin{eqnarray}}
\newcommand{\eea}{\end{eqnarray}}

\newcommand{\nn}{\nonumber}
\newcommand{\adots}{\mathinner{\mkern2mu\raise1pt\hbox{.}\mkern2mu
\raise4pt\hbox{.}\mkern2mu\raise7pt\hbox{.}\mkern1mu}}

\title{Integrable and Superintegrable Extensions \\
of the Rational Calogero-Moser Model in 3 Dimensions}

\author{Allan P. Fordy\thanks{School of Mathematics,
University of Leeds, Leeds LS2 9JT, UK. ~~E-mail: a.p.fordy@leeds.ac.uk}
$\,$ and Qing Huang\thanks{School of Mathematics, Center for Nonlinear Studies, Northwest University, Xi’an 710069,
People’s Republic of China ~~E-mail: hqing@nwu.edu.cn}
}

\begin{document}

\maketitle

\begin{abstract}
We consider a class of Hamiltonian systems in 3 degrees of freedom, with a particular type of quadratic integral and which includes the rational Calogero-Moser system as a particular case.
For the general class, we introduce separation coordinates to find the general separable (and therefore Liouville integrable) system, with two quadratic integrals.  This gives a coupling of the Calogero-Moser system with a large class of potentials, generalising the series of potentials which are separable in parabolic coordinates.  Particular cases are {\em superintegrable}, including Kepler and a resonant oscillator.

The initial calculations of the paper are concerned with the flat (Cartesian type) kinetic energy, but in Section \ref{sec:conflat-general}, we introduce a {\em conformal factor} $\varphi$ to $H$ and extend the two quadratic integrals to this case. All the previous results are generalised to this case. We then introduce some 2 and 3 dimensional symmetry algebras of the Kinetic energy (Killing vectors), which restrict the conformal factor.  This enables us to reduce our systems from 3 to 2 degrees of freedom, giving rise to many interesting systems, including both Kepler type and H\'enon-Heiles type potentials on a Darboux-Koenigs $D_2$ background.
\end{abstract}



\section{Introduction}

The Calogero-Moser system is the archetypal integrable, many-body problem, both classical and quantum \cite{71-3, 75-3}, and is known to be maximally superintegrable \cite{83-11,96-5}.  These authors derive the first integrals from the Lax matrix, but here we wish to use methods which can be used, regardless of whether or not a Lax matrix is known.  In this paper we only consider the case of 3 degrees of freedom.

The Calogero-Moser system in 3 degrees of freedom is
\begin{subequations}
\be \label{CM-Hamiltonian}
H = \frac{1}{2} (p_1^2+p_2^2+p_3^2) + g^2 \left(\frac{1}{(q_1-q_2)^2}+\frac{1}{(q_1-q_3)^2}+\frac{1}{(q_2-q_3)^2}\right),
\ee
which can be obtained from a Lax matrix $L$, as $\frac{1}{2} \mbox{tr} L^2$.  In the Lax approach, a full set of Poisson commuting integrals (for general $n$ degrees of freedom) can be written as $\frac{1}{k} \mbox{tr} L^k$.
Extending this approach, Wojciechowski \cite{83-11} found more integrals, some of which are quadratic and related to the ones given below.  Whilst these Poisson commute with $H$, they generate a {\em non-commutative} Poisson algebra, as is necessarily the case for {\em superintegrable systems}.

In this paper we particularly focus on one particular quadratic integral, given in the construction of \cite{83-11}, which can be written
\be\label{X1}
X_1 = (p_1+p_2+p_3) S-2 (q_1+q_2+q_3) H,
\ee
where $S$ is a conformal element, given by (\ref{S-def}) in the Appendix (Section \ref{sec:appendix-not}).
\end{subequations}
In formula (\ref{X1}), $H$ just denotes (\ref{CM-Hamiltonian}), but in Section \ref{X1-calc} we consider $H$ with a general potential $U$, and ask for a modification of $X_1$ (deformed with its own ``potential'' term) to be a first integral.

Indeed, in Section \ref{X1-calc} we classify the functions $U$, which allow such an integral $X_1$, by reducing the system to {\em separation variables}, $(x, y, z)$, which are a 3-dimensional extension of the {\em standard parabolic coordinates of the plane} (labelled here $(x, y)$).  Hence these potentials (in the $(x,y)$ components) are just the standard sequence (see Chapter 2 of \cite{90-16}), including the Kepler problem, the KdV case of integrable H\'enon-Heiles potential, and many more.  The $z$ component allows for the connection to the Calogero-Moser system (when re-written in the original $q_i$ coordinates).  Separation of variables gives a third quadratic function $X_2$ and the three functions $H, \, X_1,\, X_2$ play an important role throughout the paper.

To construct $(x,y,z)$, we first introduce an intermediate Cartesian coordinate system $Q_i$, in which many of these potentials take their most natural form.

Whilst Section \ref{X1-calc} discusses the general structure of Liouville integrability in the three coordinate systems, $(x,y,z),\; (Q_1,Q_2,Q_3)$ and $(q_1,q_2,q_3)$, Section 3 is concerned with special potentials in the $(Q_1,Q_2,Q_3)$ coordinates.  Whilst the general system is simply Liouville integrable, many of these special cases have additional integrals, making them {\em superintegrable}.  It's in these coordinates that such potentials as Kepler, a resonant oscillator and H\'enon-Heiles arise.

In Section \ref{sec:CMcases} the Calogero-Moser system is specifically considered with both Kepler and oscillator coupling.

Up to this point the kinetic energy has been of the standard flat, Cartesian type, but in Section \ref{sec:conflat-general}, we introduce a {\em conformal factor} $\varphi$ to the entire separable Hamiltonian $H$ (in $(x,y,z)$ coordinates), requiring it to continue to be separable in these coordinates.  This modifies the definition of $X_1$ and $X_2$, but these three functions remain in involution.  The remaining sections are concerned with this extension, mainly in terms of the coordinates $Q_i$.

Section \ref{sec:SpecPotsQP-phi} extends the cases of Section \ref{sec:SpecPotsQP} to conformally flat metrics.  Whilst the restriction of the potential is allowed without any further constraints on the conformal factor $\varphi$, demanding \underline{additional first integrals}, as in the superintegrable cases, \underline{does} force constraints on $\varphi$.  In this section we see extensions of Kepler, resonant oscillator and H\'enon-Heiles potentials.  In particular, one of the extensions to the previous resonant oscillator is a {\em superintegrable}, 3D extension of the Darboux-Koenigs $D_1$ kinetic energy with potential (see Section \ref{sec:U2-Q-phi-F3}).

In Section \ref{sec:Symms-phi} we introduce some 2 and 3 dimensional symmetry algebras (linear functions of momenta, which \underline{commute} with the \underline{kinetic energy} of $H$), discussed in detail in \cite{f20-2,f21-1}.  The existence of such symmetries restricts the conformal factor $\varphi$.  We can then restict the \underline{entire} Hamiltonian with just \underline{one} of these symmetries, after which we can choose coordinates in which the Hamiltonian can be interpreted as a 2 dimensional reduction.

Many interesting reduced systems arise in this way.  For example the Hamiltonian (\ref{keplerP1J1-Huv}) is a Kepler problem on a Darboux-Koenigs $D_2$ background.  It is superintegrable, with two independent quadratic integrals.  Similarly, the Hamiltonian (\ref{U2U3Huv}) is of generalised H\'enon-Heiles type on a Darboux-Koenigs $D_2$ background.  It has one quadratic integral, so is Liouville integrable (but not superintegrable).  The Hamiltonian (\ref{so3-H-J2uv}) is interesting in the context of Darboux-Koenigs theory; it has a Darboux-Koenigs $D_3$ kinetic energy, but the given potential only allows \underline{one quadratic} integral.  However, there is an independent \underline{quartic} integral, so the system is maximally superintegrable and outside the class considered in \cite{03-11}.

In Section \ref{sec:conf-KepCM}, we present the Kepler-Calogero-Moser system with a particular conformally flat metric.

Commonly used notations are defined in the Appendix (Section \ref{sec:appendix-not}).

\section{The General Potential Compatible with $X_1$: Flat Metric}\label{X1-calc}

We now consider the two functions
\be\label{HX1}
H = \frac{1}{2} (p_1^2+p_2^2+p_3^2)+ U(q_1,q_2,q_3), \quad X_1 = (p_1+p_2+p_3) S -2 (q_1+q_2+q_3)  H +V_1(q_1,q_2,q_3),
\ee
with $X_1$ an extension of (\ref{X1}).  We have already seen that these functions commute for the special choices (\ref{CM-Hamiltonian}) and (\ref{X1}).  We now seek conditions on the general $U$ and $V_1$ of (\ref{HX1}), such that $\{H,X_1\}=0$.

Since the leading parts of these functions are {\em quadratic} in momenta, we can use {\em separation of variables}, which corresponds to {\em simultaneous diagonalisation} of the defining quadratic forms by a canonical transformation.  We do this in two steps.  The intermediate coordinate system is not only convenient, but of interest in itself.

\subsection{Intermediate Transformation to $(Q_i,P_i)$}\label{sec:QP-coords}

We first perform an {\em orthogonal transformation}, with $Q_1$ the centre of mass coordinate:
\be\label{Qi}
Q_1=\frac{1}{\sqrt{3}} (q_1+q_2+q_3),\quad Q_2 = \frac{1}{\sqrt{2}} (q_1-q_2), \quad Q_3 = \frac{1}{\sqrt{6}} (q_1+q_2-2 q_3),
\ee
in which
$$
p_1+p_2+p_3=\sqrt{3} P_1, \quad p_1^2+p_2^2+p_3^2=P_1^2+P_2^2+P_3^2, \quad S = Q_1P_1+Q_2P_2+Q_3P_3.
$$
Therefore (up to an overall factor of $\sqrt{3}$ dropped from $X_1$), we obtain
\begin{subequations}
\be\label{HX1Q}
H = \frac{1}{2} (P_1^2+P_2^2+P_3^2)+ \bar U(Q_1,Q_2,Q_3), \quad X_1 = P_1 S -2 Q_1 H +\bar V_1(Q_1,Q_2,Q_3),
\ee
where we see that $X_1$ can be written in the same form as (\ref{HX1}).

When written explicitly, the quadratic part of $X_1$ is now simpler:
\be \label{X1Qexp}
X_1^0 =   P_1 (Q_2 P_2+Q_3 P_3) - Q_1 (P_2^2+P_3^2).
\ee
\end{subequations}
The eigenvectors of the corresponding quadratic form are each defined up to an overall {\em functional} coefficient, which can be used to write it as the gradient of a new coordinate function.

\subsection{Separation of Variables}

The details of the calculation are unimportant, but lead to the change of coordinates:
\be \label{sep-coord-xyz}
x = \sqrt{Q_1^2+Q_2^2+Q_3^2} + Q_1,\quad    y = \sqrt{Q_1^2+Q_2^2+Q_3^2} - Q_1,\quad  z = \frac{Q_3}{Q_2},
\ee
which reduce to the standard two dimensional parabolic coordinates when $Q_3=0$.  In these coordinates, the kinetic energy $H^0 = \frac{1}{2} (P_1^2+P_2^2+P_3^2)$ takes the form
\begin{subequations}
\be \label{KE-xyz}
H^0 = \frac{2 x p_x^2+2 y p_y^2}{x+y} +\frac{(1+z^2)^2 p_z^2}{2 x y},
\ee
with the pure $x-y$ part being the standard 2D parabolic coordinates case.

We can then add the standard separable potentials to (\ref{KE-xyz}):
\be \label{H-xyz}
H =   \frac{2 x p_x^2+2 y p_y^2+A_1(x)+A_2(y)}{x+y} +\frac{(1+z^2)^2 p_z^2 + 2(1+z^2) B(z)}{2 x y},
\ee
without destroying commutativity.
\br
We could have defined $z$ as $\arctan \left(\frac{Q_3}{Q_2}\right)$ to make the $z$ part just $\frac{p_z^2}{2 x y}$, but our main interest is in the $Q_i$ and $q_i$ coordinates, so such refinements are unnecessary. The definition of $B(z)$ incorporates the factor $(1+z^2)$ to simplify the expression in the $Q_i$ coordinates.
\er

Separation of variables then leads to 3 {\em quadratic} commuting integrals:
\be \label{H=E}
H=E \quad\Rightarrow\quad   2 x p_x^2+A_1(x) +\frac{\gamma}{2x}- E x+ 2 y p_y^2+A_2(y) +\frac{\gamma}{2y}- E y = 0,
\ee
where $\gamma$ is defined by
\be \label{gamma}
X_2 = (1+z^2)^2 p_z^2+2(1+z^2) B(z) = \gamma.
\ee
Defining $\alpha = 2 x p_x^2+A_1(x) +\frac{\gamma}{2x}- E x$, we can eliminate $E$ to obtain
\be \label{alpha}
\alpha = \frac{2 x y (p_x^2-p_y^2)+ y A_1(x)-x A_2(y)}{x+y} +\frac{\gamma (y-x)}{2 x y},
\ee
\end{subequations}
with $\gamma$ given by (\ref{gamma}).  Applying the transformation (\ref{sep-coord-xyz}) to $X_1$ of (\ref{HX1Q}), we see that $\alpha = X_1$.

\br[Jacobi's Theorem]
As guaranteed by Jacobi's Theorem, separability has led to a third quadratic first integral $X_2$, and that $H,\, X_1$ and $X_2$ are in involution.
\er

\subsection{Returning to $(Q_i,P_i)$ Coordinates}

Returning to these coordinates, we have
\begin{subequations}\label{HX1X2Q}
\bea
H &=& \frac{1}{2} (P_1^2+P_2^2+P_3^2)+ \frac{A_1\left(x\right)+A_2\left(y\right)}{2 R}  + \frac{B\left(\frac{Q_3}{Q_2}\right)}{Q_2^2},    \label{HQ}  \\
X_1 &=& P_1 S -2 Q_1 H +\frac{x A_1(x) - y A_2(y)}{2 R},     \label{X1Q}  \\
X_2 &=& (Q_2 P_3-Q_3 P_2)^2 +2 (Q_2^2+Q_3^2)\, \frac{B\left(\frac{Q_3}{Q_2}\right)}{Q_2^2},    \label{GQ}
\eea
where $x= R+Q_1,\; y = R-Q_1$, with $R = \sqrt{Q_1^2+Q_2^2+Q_3^2}$.
\end{subequations}

\subsection{Returning to $(q_i,p_i)$ Coordinates}

We can finally return to the original coordinates:
\begin{subequations}
\bea
H &=& \frac{1}{2} (p_1^2+p_2^2+p_3^2)+ \frac{A_1(x)+A_2(y)}{2\sqrt{q_1^2+q_2^2+q_3^2}}  + \frac{W\left(\frac{q_2-q_3}{q_1-q_2}\right)}{(q_1-q_2)^2},    \label{Hq}  \\
X_1 &=& (p_1+p_2+p_3) S -2 (q_1+q_2+q_3) H +\frac{x A_1(x) - y A_2(y)}{2 R},     \label{X1q}  \\
X_2 &=& (J_1+J_2+J_3)^2 +4 (q_1^2+q_2^2+q_3^2-q_1q_2-q_2q_3-q_3q_1)\,  \frac{W\left(\frac{q_2-q_3}{q_1-q_2}\right)}{(q_1-q_2)^2},    \label{Gq}
\eea
where {\small
$$
x= \sqrt{q_1^2+q_2^2+q_3^2} + \frac{q_1+q_2+q_3}{\sqrt{3}},\;\; y = \sqrt{q_1^2+q_2^2+q_3^2}-\frac{q_1+q_2+q_3}{\sqrt{3}}, \;\;   W\left(\frac{q_2-q_3}{q_1-q_2}\right)= 2 B\left(\frac{q_1+q_2-2 q_3}{\sqrt{3} (q_1-q_2)}\right),
$$   }
and $J_i$ are given in (\ref{piJi}).
\end{subequations}

\section{Some Specific Potentials in $(Q_i,P_i)$ Coordinates}\label{sec:SpecPotsQP}

Since the coordinates of (\ref{H-xyz}) are an extension of the 2 dimensional parabolic coordinates, the $(x,y)$ part of the potential contains several well known, important potentials, such as Kepler and one of the H\'enon-Heiles cases, which are now extended to 3 dimensions.  To appreciate this it is best to represent them in the $(Q_i,P_i)$ coordinates.  The $z$ part of the potential (\ref{H-xyz}) adds further extensions to these known potentials in the $(Q_i,P_i)$ coordinates, but is most interesting in the $(q_i,p_i)$ coordinates, since here it includes the 3 component Calogero-Moser potential as a special case.  All of these cases are completely integrable, but some have additional integrals, making them superintegrable.

\subsection{Special Choices of the Functions $A_i(Q_1,Q_2,Q_3)$}

Clearly, by setting $A_1(x)=A_2(y)=-\mu$, the potential of (\ref{HQ}) includes a Kepler part
\begin{subequations}
\be \label{U0}
H = \frac{1}{2} (P_1^2+P_2^2+P_3^2) -\, \frac{\mu}{\sqrt{Q_1^2+Q_2^2+Q_3^2}}  + \frac{B\left(\frac{Q_3}{Q_2}\right)}{Q_2^2}.
\ee
The ``$A$'' part of the potential in (\ref{HQ}) also includes an infinite family of {\em polynomial} potentials, which directly generalise the well known 2 dimensional case (see, for example, \cite{90-16}):
\be\label{Un}
U_n = k\, \frac{\left(\sqrt{Q_1^2+Q_2^2+Q_3^2}+Q_1\right)^{n+1}+ (-1)^n \left(\sqrt{Q_1^2+Q_2^2+Q_3^2}-Q_1\right)^{n+1}}{2 \sqrt{Q_1^2+Q_2^2+Q_3^2}},
\ee
with
\bea
&&  U_1 = 2 k Q_1,\quad U_2 = k (4 Q_1^2+Q_2^2+Q_3^2), \quad U_3 = 4 k Q_1 (2 Q_1^2+Q_2^2+Q_3^2), \nn\\[2mm]
&&    U_4 = k (16 Q_1^4+ 12 Q_1^2 (Q_2^2+Q_3^2) + (Q_2^2+Q_3^2)^2).   \label{Un14}
\eea
\end{subequations}
Hamiltonian (\ref{HQ}), with $U_2$ gives case 6 of Table II of \cite{90-22}.  Hamiltonian (\ref{HQ}), with a combination of $U_2$ and $U_3$ gives a 3 dimensional version of the H\'enon-Heiles system, with the addition of the $\frac{B\left(\frac{Q_3}{Q_2}\right)}{Q_2^2}$ term, which generalises the usual $\frac{c}{Q_2^2}$ term, which appears in the 2 dimensional case.  Hamiltonian (\ref{HQ}), with $U_4$ gives a generalisation of case 4 of Table 1 of \cite{86-8}.

\subsection{The Case $A_i=0$}\label{sec:Ai0-Q}

This is a well known case \cite{00-3,06-8,07-10,08-8} with 4 independent integrals for general $B\left(\frac{Q_3}{Q_2}\right)$.  Special cases of this (including the Calogero-Moser case) allow an algebra with rank 5 (which is maximal in 3 degrees of freedom).

As well as $H, X_1, X_2$, we also have $P_1$ and $\Omega$ (an extension of $J^2$, given in the Appendix):
\begin{subequations}
\bea
H &=& \frac{1}{2} (P_1^2+P_2^2+P_3^2) + \frac{B\left(\frac{Q_3}{Q_2}\right)}{Q_2^2},  \label{HA0}  \\
X_1 &=&  P_1 S - 2 Q_1 H,   \label{X1A0}  \\
X_2 &=& (Q_2 P_3-Q_3 P_2)^2 +2 (Q_2^2+Q_3^2)\, \frac{B\left(\frac{Q_3}{Q_2}\right)}{Q_2^2},    \label{X2A0}  \\
\Omega &=& (Q_1 P_2-Q_2 P_1)^2 +(Q_2 P_3-Q_3 P_2)^2 +(Q_3 P_1-Q_1 P_3)^2 + 2 (Q_1^2+Q_2^2+Q_3^2) \, \frac{B\left(\frac{Q_3}{Q_2}\right)}{Q_2^2}. \nn\\
    &=&   2 (Q_1^2+Q_2^2+Q_3^2) H -  (Q_1 P_1+Q_2 P_2+Q_3 P_3)^2. \label{OmA0}
\eea
The non-zero Poisson brackets are
\be \label{Ai0-Q-pbs}
\{P_1,X_1\}=2 H - P_1^2,\;\;\{P_1,\Omega\}=2 X_1,\;\; \{X_1,\Omega\}=-2 P_1 \Omega,
\ee
and these 5 functions satisfy the constraint
\be \label{Ai0-Q-constraint}
X_1^2 = 2 (\Omega-X_2) H -P_1^2 \Omega.
\ee
\end{subequations}

\subsection{The Kepler Case}\label{sec:kepler-Q}

As well as $H, X_1, X_2$, we also have $\Omega$ (an extension of (\ref{Jcas})):
\begin{subequations}\label{HXi:kepler-Q}
\bea
H &=& \frac{1}{2} (P_1^2+P_2^2+P_3^2) -\, \frac{\mu}{\sqrt{Q_1^2+Q_2^2+Q_3^2}}  + \frac{B\left(\frac{Q_3}{Q_2}\right)}{Q_2^2},  \label{HU0}  \\
X_1 &=&  P_1 S - 2 Q_1 H -\frac{\mu Q_1}{\sqrt{Q_1^2+Q_2^2+Q_3^2}},   \label{X1U0}  \\
X_2 &=& (Q_2 P_3-Q_3 P_2)^2 +2 (Q_2^2+Q_3^2)\, \frac{B\left(\frac{Q_3}{Q_2}\right)}{Q_2^2},    \label{X2U0}  \\
\Omega &=& (Q_1 P_2-Q_2 P_1)^2 +(Q_2 P_3-Q_3 P_2)^2 +(Q_3 P_1-Q_1 P_3)^2 + 2 (Q_1^2+Q_2^2+Q_3^2) \, \frac{B\left(\frac{Q_3}{Q_2}\right)}{Q_2^2}. \nn\\
    &=&   2 (Q_1^2+Q_2^2+Q_3^2) H -  (Q_1 P_1+Q_2 P_2+Q_3 P_3)^2 + 2 \mu  \sqrt{Q_1^2+Q_2^2+Q_3^2} . \label{OmU0}
\eea
As well as $\{X_1,X_2\}=\{X_2,\Omega\}=0$, we have $\{X_1,\Omega\}=-2 X_3$, where
\be
X_3 = P_1 \Omega + \frac{\mu}{\sqrt{Q_1^2+Q_2^2+Q_3^2}}\, \left(Q_2 (Q_1 P_2-Q_2 P_1)+ Q_3 (Q_1 P_3-Q_3 P_1)\right),  \label{X3U0}
\ee
with $\{\Omega,X_3\} = -2 X_1 \Omega$.  We also have
$$
\{X_1,X_3\} = 2 (X_2-2\Omega) H + X_1^2 - \mu^2,\quad  X_3^2 = \Omega (2 \Omega H - 2 X_2 H - X_1^2)+\mu^2 (\Omega-X_2).
$$
\end{subequations}

\subsection{The Oscillator Potential $U_2$}\label{sec:U2-Q}

We start with the 3 integrals related to the separation of variables:
\begin{subequations}
\bea
H &=& \frac{1}{2} (P_1^2+P_2^2+P_3^2) +\frac{1}{2} \omega^2\, (4 Q_1^2+Q_2^2+Q_3^2)  + \frac{B\left(\frac{Q_3}{Q_2}\right)}{Q_2^2},  \label{HU2}  \\
X_1 &=&  P_1 S - 2 Q_1 H + 2\omega^2\, Q_1 (2 Q_1^2+Q_2^2+Q_3^2), \label{X1U2}  \\
X_2 &=& (Q_2 P_3-Q_3 P_2)^2 +2 (Q_2^2+Q_3^2)\, \frac{B\left(\frac{Q_3}{Q_2}\right)}{Q_2^2}.    \label{GU2}
\eea
Since $B$ is independent of $Q_1$, we see that we have an additional integral:
\be \label{F1U2}
F_1 = P_1^2 + 4 \omega^2 Q_1^2 .
\ee
We can then define $F_2$ by
\be \label{F2U2}
\{X_1,F_1\} = -2 F_2 \quad\Rightarrow\quad  F_2 = P_1 (2 H-P_1^2)+2 \omega^2 (2 Q_1 (Q_2 P_2+Q_3 P_3)-P_1  (2 Q_1^2+Q_2^2+Q_3^2)).
\ee
The remaining nonzero Poisson brackets are
\be\label{U2-pbs}
\{X_1,F_2\} = (F_1-2 H) (2 H-3 F_1) +4 \omega^2 X_2,\;\; \{F_1,F_2\} = -8 \omega^2 X_1.
\ee
These integrals have rank $4$, satisfying
\be \label{U2-F2cond}
F_2^2 = F_1 (F_1-2 H)^2-4 \omega^2 (F_1 X_2+X_1^2).
\ee
\end{subequations}

\subsubsection{Relation to Resonant Oscillator}\label{sec:ResOs}

When $B=0$, this is just a 3 dimensional {\em resonant oscillator}, which can be thought of as a combination of 2 component oscillators, with resonances $(2,1),\, (2,1)$ and $(1,1)$.

The 2 dimensional oscillator with resonances $(2,1)$ has quadratic and cubic integrals.  We can write $X_1$ and $F_2$ as the sum of integrals in the $(Q_1,Q_2)$ and $(Q_1,Q_3)$ spaces, together with an extra piece involving $B$:
\begin{subequations}
\bea
X_1 &=& (-Q_1 P_2^2+Q_2 P_1 P_2+\omega^2 Q_1 Q_2^2) + (-Q_1 P_3^2+Q_3 P_1 P_3+\omega^2 Q_1 Q_3^2) -2 Q_1 \, \frac{B\left(\frac{Q_3}{Q_2}\right)}{Q_2^2},  \label{U2-X1}\\
F_2 &=& (P_1 P_2^2-\omega^2 Q_2^2 P_1+ 4 \omega^2 Q_1 Q_2 P_2) + (P_1 P_3^2-\omega^2 Q_3^2 P_1+ 4 \omega^2 Q_1 Q_3 P_3) +2 P_1 \, \frac{B\left(\frac{Q_3}{Q_2}\right)}{Q_2^2}, \label{U2-F2}
\eea
The integral $X_2$ is itself related to the $(1,1)$ resonance, as is another {\em quadratic} integral:
\be \label{U2-F3V2}
F_3^{(2)} = P_2^2 +\omega^2 Q_2^2 + V_3^{(2)}(Q_2),
\ee
where the additional term $V_3^{(2)}(Q_2)$ is to be determined. The condition $\{H,F_3^{(2)}\}=0$ leads to
\be \label{U2-F3B}
B=Q_2^2 \left( \frac{b_1}{Q_2^2}+\frac{b_2}{Q_3^2} \right),\quad V_3^{(2)} =\frac{2b_1}{Q_2^2},
\ee
giving (\ref{HU2}) as the last case in Table I of \cite{90-22}.  We see that there is an additional integral
\be \label{U2-F3V3}
F_3^{(3)} = P_3^2 +\omega^2 Q_3^2 + \frac{2b_2}{Q_3^2}.
\ee
We also have:
\be\label{X12}
X_1 = P_1 S -2 Q_1 H +2 \omega^2 Q_1 (2 Q_1^2+Q_2^2+Q_3^2),\quad X_2 = (Q_2P_3-Q_3P_2)^2 +2 (Q_2^2+Q_3^2) \left(\frac{b_1}{Q_2^2} +\frac{b_2 }{Q_3^2}\right),
\ee
\end{subequations}
with $F_1,\, F_3^{(2)},\, F_3^{(3)},\, X_1,\, X_2$ being rank 5.  They generate a 10 dimensional Poisson algebra, with 5 polynomial constraints, all of which can be derived as a restriction of the conformally flat case, discussed in Sections \ref{sec:U2-Q-phi-F3} and \ref{sec:appendix-pbs}.  This algebra is discussed in Section \ref{sec:appendix-pbs-flat} of the Appendix.

\subsection{The Generalised H\'enon-Heiles Potential $U_3$}\label{sec:U3-Q}

We have the three functions in involution:
\begin{subequations}
\bea
H &=& \frac{1}{2} (P_1^2+P_2^2+P_3^2) +\frac{1}{2} \omega^2\, (4 Q_1^2+Q_2^2+Q_3^2) + 4 k Q_1 (2 Q_1^2+Q_2^2+Q_3^2) + \frac{B\left(\frac{Q_3}{Q_2}\right)}{Q_2^2},  \label{HU3}  \\
X_1 &=&  P_1 S - 2 Q_1 H + 2\omega^2\, Q_1 (2 Q_1^2+Q_2^2+Q_3^2)  \nn\\
   &&    \hspace{6cm}  + k (16 Q_1^4+ 12 Q_1^2 (Q_2^2+Q_3^2) + (Q_2^2+Q_3^2)^2)  \nn\\
   &=& P_3J_2-P_2 J_3+\omega^2 Q_1 (Q_2^2+Q_3^2) + k (Q_2^2+Q_3^2)^2(4 Q_1^2+Q_2^2+Q_3^2) -2 Q_1\, \frac{B\left(\frac{Q_3}{Q_2}\right)}{Q_2^2},   \label{X1U3} \\
X_2 &=& (Q_2 P_3-Q_3 P_2)^2 +2 (Q_2^2+Q_3^2)\, \frac{B\left(\frac{Q_3}{Q_2}\right)}{Q_2^2},    \label{GU3}
\eea
so it is completely integrable.  This is a 3 degrees of freedom generalisation of the standard ``KdV case'' of integrable H\'enon-Heiles models (see \cite{f91-1}). The $B-$term generalises the usual $\frac{c}{Q_2^2}$ term.
\end{subequations}

\section{The $(q_i,p_i)$ Coordinates and the Calogero-Moser Potential}\label{sec:CMcases}

We now consider these systems in the original coordinates $(q_i,p_i)$.  Our main interest is the Calogero-Moser Potential with $B(z)$ (from (\ref{HQ})) and $W(\zeta)$ (from (\ref{Hq})) taking the specific forms
\be \label{BW-CM}
B(z) = \frac{9 g^2 (1+z^2)^2}{2 (1-3 z^2)^2},\quad  W(\zeta) = g^2 \left(1 +\frac{1}{\zeta^2} +\frac{1}{(1+\zeta)^2}\right).
\ee

\subsection{Kepler-Calogero-Moser System}\label{sec:KCM}

Rewriting the Kepler case of Section \ref{sec:kepler-Q}, we find the Kepler-Calogero-Moser System \cite{17-5}:
\begin{subequations}
\bea
H &=&  \frac{1}{2} (p_1^2+p_2^2+p_3^2) + g^2 \left(\frac{1}{(q_1-q_2)^2}+\frac{1}{(q_1-q_3)^2}+\frac{1}{(q_2-q_3)^2}\right) -\frac{\mu}{\sqrt{q_1^2+q_2^2+q_3^2}},  \label{H-CMK} \\
X_1 &=& (p_1+p_2+p_3) S-2 (q_1+q_2+q_3) H - \frac{\mu (q_1+q_2+q_3)}{\sqrt{q_1^2+q_2^2+q_3^2}},  \label{X1-CMK} \\
X_2 &=& (J_1+J_2+J_3)^2  \nn\\
&&  +4 g^2 (q_1^2+q_2^2+q_3^2-q_1q_2-q_2q_3-q_3q_1) \left(\frac{1}{(q_1-q_2)^2}+\frac{1}{(q_1-q_3)^2}+\frac{1}{(q_2-q_3)^2}\right), \label{X2-CMK}  \\
 \Omega &=& 2 (q_1^2+q_2^2+q_3^2) H-S^2 +2 \mu \sqrt{q_1^2+q_2^2+q_3^2} .    \label{Om-CMK}
\eea

As well as $\{H,X_i\}=0$, we have a further {\em cubic} integral $X_3$, defined by $\{X_1,\Omega\}= - 2X_3$, where
\bea
X_3 &=& (p_1+p_2+p_3) \Omega    -\frac{\mu}{\sqrt{q_1^2+q_2^2+q_3^2}} \, ((q_2^2+q_3^2-q_1(q_2+q_3))p_1  \nn\\
     &&  \qquad  +(q_3^2+q_1^2-q_2(q_3+q_1))p_2+(q_1^2+q_2^2-q_3(q_1+q_2))p_3).  \label{X3-CMK}
\eea
The remaining nonzero Poisson relations are listed below
\be\label{XiXj}
 \{X_1,X_3\}= 2 (X_2-6 \Omega) H + X_1^2-3\mu^2,\quad  \{\Omega,X_3\}= -2 X_1 \Omega.
\ee
The functions $H,X_1,X_2,\Omega,X_3$ have rank 4 and satisfy
\be \label{X3^2=}
X_3^2 = \left(2(3 \Omega- X_2) H - X_1^2+3\mu^2\right)\Omega-\mu^2 X_2.
\ee
\end{subequations}

\subsection{A Calogero-Moser System Coupled with the Potential $U_2$}

Rewriting the case of Equation (\ref{HU2}), we obtain
\begin{subequations}
\bea
H &=&  \frac{1}{2} (p_1^2+p_2^2+p_3^2) + g^2 \left(\frac{1}{(q_1-q_2)^2}+\frac{1}{(q_1-q_3)^2}+\frac{1}{(q_2-q_3)^2}\right)\nn\\
&&   \hspace{5cm}  +\omega^2 (q_1^2+q_2^2+q_3^2+q_1q_2+q_2q_3+q_3q_1),  \label{H-CMOsc} \\
X_1 &=& (p_1+p_2+p_3) S -2 (q_1+q_2+q_3) H \nn\\
    &&  \hspace{3cm}   + \frac{4}{3} \omega^2 (q_1+q_2+q_3) (2q_1^2+2q_2^2+2q_3^2+q_1q_2+q_2q_3+q_3q_1) ,  \label{X1-CMOsc} \\
X_2 &=&  (J_1+J_2+J_3)^2   \nn\\
    &&   \qquad +4 g^2\, (q_1^2+q_2^2+q_3^2-q_1q_2-q_2q_3-q_3q_1) \left(\frac{1}{(q_1-q_2)^2}+\frac{1}{(q_1-q_3)^2}+\frac{1}{(q_2-q_3)^2}\right), \label{G-CMOsc} \\
 F_1 &=& (p_1+p_2+p_3)^2+ 4 \omega^2 (q_1+q_2+q_3)^2.    \label{F1-CMOsc}
\eea
\end{subequations}

We then have $\{X_1,F_1\} = -4 F_2$, where
\begin{subequations}
\bea
F_2 &=& F_2^{(1)} -3 F_2^{(2)} +3 \omega^2(q_1+q_2+q_3) (q_1p_1+q_2p_2+q_3p_3)  \nn\\
    &&   \qquad   -3 \omega^2 \left((q_2^2+q_3^2+q_2q_3) p_1+(q_1^2+q_3^2+q_1q_3) p_2+(q_1^2+q_2^2+q_1q_2) p_3\right),   \label{F2-CMU2}
\eea
with
\bea
F_2^{(1)} &=& p_1^3+p_2^3+p_3^3 +3 g^2 \left(\frac{p_1+p_2}{(q_1-q_2)^2}+\frac{p_1+p_3}{(q_1-q_3)^2}+\frac{p_2+p_3}{(q_2-q_3)^2}\right) , \label{F21-CMU2}  \\
F_2^{(2)} &=&  p_1 p_2 p_3 - g^2 \left(\frac{p_3}{(q_1-q_2)^2}+\frac{p_2}{(q_1-q_3)^2}+\frac{p_1}{(q_2-q_3)^2}\right).  \label{F22-CMU2}
\eea
We write it like this because, when $\omega=0$, both $F_2^{(1)}$ and $F_2^{(2)}$ are first integrals.

\medskip
The remaining nonzero brackets are
\be \label{PBs-CMU2}
\{F_1,F_2\} = -36 \omega^2 X_1, \quad  \{X_1,F_2\} = -\frac{3}{2} (6 H-F_1) (2 H-F_1) +6 \omega^2 X_2.
\ee
This algebra has rank 4, since
\be \label{F2square-CMU2}
F_2^2 = \frac{1}{4} F_1 (F_1-6 H)^2 -3 \omega^2 (F_1 X_2 + 3 X_1^2).
\ee

\br
Writing the oscillator term as
\be \label{isotropic-plus}
\omega^2 (q_1^2+q_2^2+q_3^2+q_1q_2+q_2q_3+q_3q_1) = 2 \omega^2 (q_1^2+q_2^2+q_3^2)-\frac{1}{2}\, \omega^2 ((q_1-q_2)^2+(q_2-q_3)^2+(q_3-q_1)^2),
\ee
it can be interpreted as a modification of the ``nearest neighbour interaction'', together with an external isotropic oscillator.
\er
\end{subequations}

\section{The Conformally Flat Case}\label{sec:conflat-general}

In this and the following sections, we extend our results to the conformally flat case. We \underline{start} in the separation coordinates, $(x,y,z)$, adding a {\em conformal factor} to the entire separable Hamiltonian (\ref{H-xyz}), but requiring that the resulting Hamiltonian is still separable in these coordinates.  The functions $X_1,\, X_2$ are thus extended to this case.

In Section \ref{sec:conflat-QP}, we transform these functions into the $(Q_i,P_i)$ coordinates of (\ref{Qi}) and then (in Section \ref{sec:SpecPotsQP-phi}) extend the results of Section \ref{sec:SpecPotsQP}.

In Section \ref{sec:Symms-phi} we consider restrictions of the conformal factor by imposing various 2 and 3 dimensional symmetry algebras on the kinetic energy.

\subsection{In the $(x,y,z)$ Coordinates}\label{sec:conflat-xyz}

We add a {\em conformal factor} to the entire separable Hamiltonian (\ref{H-xyz})
\begin{subequations}\label{HX1X2phi-xyz}
\be \label{Hphi-xyz}
H=\varphi(x,y,z)\left(\frac{2x p_x^2+2y p_y^2+A_1(x)+A_2(y)}{x+y}+\frac{(1+z^2)^2 p_z^2+2(1+z^2)B(z)}{2xy}\right),
\ee
demanding that $H=E$ can still be separated.  This gives that $\varphi$ must have the form
\be \label{phi-xyz}
 \varphi=\frac{xy(x+y)}{xy(\varphi_1(x)+\varphi_2(y))+(x+y)\varphi_3(z)},
\ee
and repeating the construction of $X_1$ and $X_2$, we see they are modified as follows:
\bea
X_1 &=& \frac{2(x^2p_x^2-y^2p_y^2)}{x+y}+\frac{y\varphi_2(y)-x\varphi_1(x)}{x+y}H+\frac{xA_1(x)-yA_2(y)}{x+y},   \label{X1phi-xyz}\\
X_2 &=& (1+z^2)^2 p_z^2+2(1+z^2)B(z)-2\varphi_3(z)H,    \label{X2phi-xyz}
\eea
\end{subequations}
with $H,\, X_1$ and $X_2$ still in involution.

\subsection{In the $(Q_i,P_i)$ Coordinates}\label{sec:conflat-QP}

Rewriting this in the $(Q_i,P_i)$ coordinates, the system (\ref{HX1X2phi-xyz}) takes the form
\begin{subequations}\label{HX1X2phi-QP}
\bea
  H &=& \varphi\left(\frac12\left(P_1^2+P_2^2+P_3^2\right)+\frac{A_1(x)+A_2(y)}{2R}+\frac1{Q_2^2}B\left(\frac{Q_3}{Q_2}\right)\right),   \label{Hphi-QP}\\[2mm]
  X_1 &=& P_1S+\frac{y\varphi_2(y)-x\varphi_1(x)}{2R}H+\frac{xA_1(x)-yA_2(y)}{2R},   \label{X1phi-QP}\\[2mm]
  X_2 &=& (Q_2P_3-Q_3P_2)^2-2\varphi_3\left(\frac{Q_3}{Q_2}\right)H+\frac{2(Q_2^2+Q_3^2)}{Q_2^2}B\left(\frac{Q_3}{Q_2}\right),   \label{X2phi-QP}
\eea
where
\be \label{phi-QP}
 \varphi=\frac{2(Q_2^2+Q_3^2)R}{(Q_2^2+Q_3^2)(\varphi_1(x)+\varphi_2(y))+2R\varphi_3\left(\frac{Q_3}{Q_2}\right)},
\ee
with $x=R+Q_1$, $y=R-Q_1$ and $R=\sqrt{Q_1^2+Q_2^2+Q_3^2}$.
\end{subequations}

\medskip
{\em When $\varphi_1(x)=x,\ \varphi_2(y)=y,\ \varphi_3\left(\frac{Q_3}{Q_2}\right)=0$, this system reduces to (\ref{HX1X2Q}).}
In this way, we extend the {\em completely integrable} system (\ref{HX1X2Q}) to a conformally flat case.

\section{Some Specific Potentials in the Conformally Flat Case}\label{sec:SpecPotsQP-phi}

Here we extend the cases of Section \ref{sec:SpecPotsQP} to conformally flat metrics.  Whilst the restriction of the functions $A_i$ is allowed without any further constraints on the conformal factor $\varphi$, demanding \underline{additional first integrals} (such as $\Omega$ in the Kepler case) \underline{does} force constraints on $\varphi$.

\subsection{The Case $A_i=0$}\label{sec:Ai0-Q-phi}

Setting $A_i=0$ (as in Section \ref{sec:Ai0-Q}) does not change the result that the functions (\ref{HX1X2phi-QP}) are in involution, so is no constraint on the function $\varphi$.  However, asking for an extension of the function (\ref{OmA0}) to commute with $H$ \underline{does} restrict $\varphi$.  We omit the details, but requiring $\{H,\Omega\}=0$, where $\Omega$ has the form
\begin{subequations}\label{Ai0-Q-phi}
\be\label{Ai0-Q-Om-phi}
  \Omega=-S^2+\psi(Q_1,Q_2,Q_3)H+V(Q_1,Q_2,Q_3),
\ee
leads to
\be\label{Ai0-Q-phi-sol-i}
  \varphi_1 =k_1x+\frac{k_2}x,\quad \varphi_2=k_1y+\frac{k_3}y+k_4,\quad \psi=2k_1(Q_1^2+Q_2^2+Q_3^2)+k_4\sqrt{Q_1^2+Q_2^2+Q_3^2},\quad V=0,
\ee
giving
\be \label{Ai0-Q-phi-sol}
 \varphi=\frac{2(Q_2^2+Q_3^2)R}{\left(2k_1(Q_2^2+Q_3^2)+2\varphi_3\left(\frac{Q_3}{Q_2}\right)+k_2+k_3\right)R+k_4(Q_2^2+Q_3^2)+(k_3-k_2)Q_1}.
\ee
\end{subequations}
Imposing the additional constraint $\{H,P_1\}=0$ leads to $k_3-k_2=k_4=0$, giving
\begin{subequations}\label{Ai0-Q-phi-1}
\be\label{Ai0-Q-H-phi-1}
  H=\frac{Q_2^2+Q_3^2}{k_1(Q_2^2+Q_3^2)+\varphi_3\left(\frac{Q_3}{Q_2}\right)+k_2}\left(\frac12\left(P_1^2+P_2^2+P_3^2\right)+\frac1{Q_2^2}B\left(\frac{Q_3}{Q_2}\right)\right),
\ee
with similar modifications to $X_1, X_2$ and $\Omega$, which now satisfy
\be\label{Ai0-Q-phi-pbs}
   \{P_1,X_1\}=2k_1H-P_1^2,\quad \{P_1,\Omega\}=2X_1,\quad \{X_1,\Omega\}=-2 P_1\Omega.
\ee
These 5 functions have rank 4, satisfying the constraint
\be\label{Ai0-Q-phi-con}
  X_1^2=4k_1k_2H^2+2k_1H(\Omega-X_2)-P_1^2\Omega,
\ee
generalising (\ref{Ai0-Q-constraint}).
\end{subequations}

\br[Further Symmetry Constraint]
If we set $\varphi_3=0$, in (\ref{Ai0-Q-H-phi-1}), then the metric has an additional rotational symmetry (with respect to $J_1$) and coincides with that of (4a) in \cite{f20-2}.  If we ask for the \underline{whole} Hamiltonian to commute with this rotation, then we obtain
$$
  H=\frac{Q_2^2+Q_3^2}{k_1(Q_2^2+Q_3^2)+k_2}\left(\frac{1}{2}\left(P_1^2+P_2^2+P_3^2\right)+\frac{k_5}{Q_2^2+Q_3^2}\right),
$$
which is a reduced form of (4.6a) in \cite{f21-1}.  Such constraints are discussed further in Section \ref{sec:Symms-phi}.
\er

\subsection{The Kepler Case}\label{sec:Kepler-Q-phi}

Here we extend the integrals (\ref{HXi:kepler-Q}), to obtain
\begin{subequations}\label{HXi:kepler-Q-phi}
\bea
  H &=& \varphi\left(\frac{1}{2}\left(P_1^2+P_2^2+P_3^2\right)-\frac{\mu}{R}+\frac1{Q_2^2}B\left(\frac{Q_3}{Q_2}\right)\right),\label{H:kepler-Q-phi}\\
  X_1 &=& P_1S+\left(-2k_1Q_1+\frac{k_3}2-\frac{k_3Q_1-k_2}{2R}\right)H-\frac{\mu Q_1}{R}, \label{X1:kepler-Q-phi} \\
  X_2 &=& (Q_2P_3-Q_3P_2)^2-2\varphi_3\left(\frac{Q_3}{Q_2}\right)H+\frac{2(Q_2^2+Q_3^2)}{Q_2^2}B\left(\frac{Q_3}{Q_2}\right), \label{X2:kepler-Q-phi} \\
  \Omega &=& -S^2+\left(2k_1 R^2+k_3R\right)H+2\mu R,  \label{Om:kepler-Q-phi}
\eea
with
\be\label{phi:kepler-Q-phi}
  \varphi=\frac{2(Q_2^2+Q_3^2)R}{2\left(k_1(Q_2^2+Q_3^2)+\varphi_3\left(\frac{Q_3}{Q_2}\right)\right)R+k_3(Q_2^2+Q_3^2)+k_2Q_1}.
\ee
\end{subequations}
Again, this restriction on $\varphi$ is a result of demanding the existence of the extended $\Omega$.

\subsection{The Oscillator Potential $U_2$}\label{sec:U2-Q-phi}

Again, specifying $A_i$ to obtain $U_2$ does not change the result that the functions (\ref{HX1X2phi-QP}) are in involution, so is no constraint on the function $\varphi$.  However, asking for an extension of either (\ref{F1U2}) or (\ref{U2-F3V2}) does lead to restrictions.  Unlike the flat case, we cannot impose {\em both} of these simultaneously, without forcing $\varphi$ to be \underline{constant}, thus returning to the flat case.

\subsubsection{Extending (\ref{F1U2})}

With $F_1=P_1^2+\psi(Q_1,Q_2,Q_3)H+V(Q_1,Q_2,Q_3)$, we find $\varphi_1(x)=k_1x+\frac{k_2}x,\;\varphi_2(y)=k_1y+\frac{k_2}y$, leading to
\begin{subequations}\label{U2-Q-phi-F1}  {\small
\bea
  H &=&  \frac{Q_2^2+Q_3^2}{k_1(Q_2^2+Q_3^2)+k_2+\varphi_3\left(\frac{Q_3}{Q_2}\right)}
  \left(\frac12\left(P_1^2+P_2^2+P_3^2\right)+\frac{\omega^2}2(4Q_1^2+Q_2^2+Q_3^2)+\frac1{Q_2^2}B\left(\frac{Q_3}{Q_2}\right)\right),\label{H:U2-Q-phi-F1}\\
  X_1 &=& P_1S-2k_1Q_1H+2\omega^2 Q_1(2Q_1^2+Q_2^2+Q_3^2),  \label{X1:U2-Q-phi-F1}\\
  X_2 &=& (Q_2P_3-Q_3P_2)^2-2\varphi_3\left(\frac{Q_3}{Q_2}\right)H+\frac{2(Q_2^2+Q_3^2)}{Q_2^2}B\left(\frac{Q_3}{Q_2}\right),  \label{X2:U2-Q-phi-F1}\\
  F_1 &=& P_1^2+4\omega^2 Q_1^2.  \label{F1:U2-Q-phi-F1}
\eea   }
These four functions generate the Poisson algebra:
\be\label{pbs:U2-Q-phi}
 \{X_1,F_1\}=-2F_2,\;\; \{X_1,F_2\}=(3F_1-2k_1H)(2k_1H-F_1)+4\omega^2 (X_2-2 k_2 H),\;\; \{F_1,F_2\}=-8\omega^2 X_1,
\ee
where
\be\label{F2:U2-Q-phi-F1}
F_2=P_1(2k_1H-P_1^2)+2\omega^2(2Q_1(Q_2P_2+Q_3P_3)-(2Q_1^2+Q_2^2+Q_3^2)P_1) .
\ee
These 5 functions have rank 4, satisfying
\be\label{con:U2-Q-phi}
 F_2^2=F_1(2k_1H-F_1)^2-4\omega^2(X_1^2+X_2F_1-2 k_2 F_1 H).
\ee
\end{subequations}

\subsubsection{Extending (\ref{U2-F3V2})}\label{sec:U2-Q-phi-F3}

With $F_3^{(2)}=P_2^2+\omega^2 Q_2^2+V_3^{(2)}(Q_2)$, we find
\bea
&&  \varphi_1(x)=k_3x^2+k_1x-\frac{k_5}x,\quad \varphi_2(y)=-k_3y^2+k_1y-\frac{k_5}y,\quad \varphi_3(z)=\frac{k_6}{z^2}+k_5+k_6,\nn\\
&&  B(z)=\frac{k_2}2+\frac{k_4}{2z^2},\quad V_3^{(2)}(Q_2)=\frac{k_2}{Q_2^2},  \nn
\eea
leading to
$$
H = \frac{Q_3^2}{(2k_3Q_1+k_1)Q_3^2+k_6}\left(\frac12\left(P_1^2+P_2^2+P_3^2\right)+\frac{\omega^2}2 (4Q_1^2+Q_2^2+Q_3^2)
     +\frac{k_2}{2Q_2^2}+\frac{k_4}{2Q_3^2}\right).
$$
Unlike the flat case, we must {\em separately} require the further integral
$$
F_3^{(3)}=P_3^2+\omega^2 Q_3^2+V_3^{(3)}(Q_3), \quad\mbox{which imposes}\quad k_6=0,\; V_3^{(3)}(Q_2)=\frac{k_4}{Q_3^2},
$$
leading to a conformally flat version of the last case in Table 1 of \cite{90-22}:
\begin{subequations}\label{U2-Q-phi-F3}     {\small
\bea
  && H = \frac{1}{2k_3Q_1+k_1}\left(\frac12\left(P_1^2+P_2^2+P_3^2\right)+\frac{\omega^2}2 (4Q_1^2+Q_2^2+Q_3^2)
     +\frac{k_2}{2Q_2^2}+\frac{k_4}{2Q_3^2}\right),\label{H:U2-Q-phi-F3}\\
  &&  X_1 = P_1S-(k_3(4Q_1^2+Q_2^2+Q_3^2)+2k_1Q_1)H+2\omega^2 Q_1(2Q_1^2+Q_2^2+Q_3^2),\label{X1:U2-Q-phi-F3}\\
  &&  X_2 = (Q_2P_3-Q_3P_2)^2+ (Q_2^2+Q_3^2) \left(\frac{k_2}{Q_2^2}+\frac{k_4}{Q_3^2}\right),  \label{X2:U2-Q-phi-F3}\\
  && X_3 = P_2^2+\omega^2Q_2^2+\frac{k_2}{Q_2^2},  \quad X_4 = P_3^2+\omega^2Q_3^2+\frac{k_4}{Q_3^2}. \label{F3:U2-Q-phi-X3X4}
\eea    }
\end{subequations}
This Hamiltonian is clearly separable and this separability is directly related to the involutive triple $H, X_3, X_4$.  We also have the involutive triple $H, X_1, X_2$, which is related to separation in the $x, y, z$ coordinates.

We can add two more simple quadratic integrals:
\begin{subequations}
\be\label{U2-Q-phi-X5X6}
X_5=P_2J_3+k_3Q_2^2\, H+Q_1\left(\frac{k_2}{Q_2^2}-\omega^2Q_2^2\right),\qquad
  X_6=-P_3J_2+k_3Q_3^2\, H+Q_1\left(\frac{k_4}{Q_3^2}-\omega^2Q_3^2\right),
\ee
which satisfy $X_1+X_5+X_6=0$, so have essentially replaced $X_1$ by two simpler integrals. These are related to integrals of 2D resonant oscillators, as noted in (\ref{U2-X1}) for the flat case.

To build the Poisson algebra generated by these integrals, we need four additional cubic integrals:
\be\label{U2-Q-phi-X78910}
  X_7=\frac14\{X_2,X_3\},\quad X_8=\frac12\{X_2,X_5\},\quad X_9=\frac12\{X_3,X_5\},\quad X_{10}=\frac12\{X_4,X_6\}.
\ee
The remaining Poisson brackets between the {\em quadratic} elements $X_2,\dots, X_6$ are given by
\be\label{U2-Q-phi-quadPBs}
\{X_2,X_3+X_4\}=\{X_2,X_5+X_6\}=\{X_3,X_4\}=\{X_3,X_6\}=\{X_4,X_5\}=0,\quad \{X_5,X_6\}=X_7.
\ee
The brackets with the cubic elements are more complicated, so given in the Appendix (Section \ref{sec:appendix-pbs}).

The integrals $H,\,X_2,\, X_3,\, X_4$ and $X_5$ are functionally independent, so this system is maximally superintegrable.  The entire algebra, including $H$, is 10 dimensional, so subject to polynomial constraints, also given in the Appendix.
\end{subequations}

\br[Darboux-Koenigs $D_1$]
On the level surface $X_4=m_4$, the Hamiltonian (\ref{H:U2-Q-phi-F3}) reduces to
$$
H = \frac{1}{2k_3Q_1+k_1}\left(\frac12\left(P_1^2+P_2^2\right)+\frac{\omega^2}2 (4Q_1^2+Q_2^2)+\frac{k_2}{2Q_2^2} +\frac{m_4}{2}\right),
$$
which is the Darboux-Koenigs $D_1$ kinetic energy with a Case 1 potential in the classification of \cite{02-6}.  Two quadratic integrals which commute with $X_4$ are $X_3$ and $X_5$, which are the ones listed in  \cite{02-6}.
\er

\subsection{The Generalised H\'enon-Heiles Potential $U_3$}

Choosing $A_i$ to give the form of (\ref{HU3}), imposes no restrictions on $\varphi$.  However, imposing $\{\varphi, P_1\}=0$ (Killing vector of the metric), leads to
\begin{subequations}
\be\label{U2U3phii}
\varphi_1(x)=k_1x+\frac{k_2}x,\quad \varphi_2(y)=k_1y+\frac{k_2}y, \quad\mbox{so}\quad \varphi = \frac{Q_2^2+Q_3^2}{k_1(Q_2^2+Q_3^2)+k_2+\varphi_3\left(\frac{Q_3}{Q_2}\right)}.
\ee
Now requiring the {\em entire} Hamiltonian to commute with $J_1=Q_2P_3-Q_3P_2$, implies that $\varphi_3=\mbox{const}$ (absorbed into $k_2$) and $B(z)=\frac{k_3}{1+z^2}$, giving
{\small  \bea
 H &=& \frac{Q_2^2+Q_3^2}{k_1(Q_2^2+Q_3^2)+k_2}\left(\frac12\left(P_1^2+P_2^2+P_3^2\right)+\frac{\omega^2}2(4Q_1^2+Q_2^2+Q_3^2)
                                                                                 +4k Q_1(2Q_1^2+Q_2^2+Q_3^2)+\frac{k_3}{Q_2^2+Q_3^2}\right),  \nn\\
  &&            \label{U2U3HX1}   \\
   X_1 &=& P_1S-2k_1Q_1H+2\omega^2Q_1(2Q_1^2+Q_2^2+Q_3^2) +  k\left(16Q_1^4+12Q_1^2(Q_2^2+Q_3^2)+(Q_2^2+Q_3^2)^2\right),  \nn
\eea    }
with $X_2$ reducing to $J_1^2+2 k_3$.
\end{subequations}

In the context of \cite{f20-2,f21-1}, the corresponding metric has symmetry algebra $\left< e_1,h_4\right>$ (where $h_4 = J_1$), so we can adapt our ``universal coordinates'' to reduce this Hamiltonian to 2 degrees of freedom.  This is discussed in Section \ref{sec:Symms-P1J1}.

\section{The Conformally Flat Case with Symmetries}\label{sec:Symms-phi}

In \cite{f20-2,f21-1} we made a systematic study of subalgebras of the full conformal algebra (in 3 dimensions).  For each subalgebra in the classification, we considered the restriction of the {\em general} conformal factor $\varphi$, in order to be invariant under the action of the subalgebra.

Here we consider such restrictions of our conformal factor $\varphi$, given by (\ref{phi-QP}).  Our aim is not to give a full classification, but to focus on some of the more important examples.  As explained in \cite{f21-1}, for each symmetry algebra, there exist {\em universal coordinates}, in which the kinetic energy is in separable form.  In \cite{f21-1}, we then added separable potentials, but \underline{here} we are \underline{given} a potential by the construction.  We can, however, restrict this to be invariant with respect to \underline{one} of the symmetries.  The resulting Hamiltonian can be written in terms of the corresponding universal coordinates, making one of the coordinates ``ignorable'', so the Hamiltonian can be interpreted as a 2 dimensional reduction.  First integrals which {\em commute} with the adapted symmetry can also be reduced to this 2 dimensional space.

The formulae for $S$ and $J_i$ are given in (\ref{AppenFormulae}), in terms of the coordinates $q_i,p_i$.  Here, of course, we need the equivalent formulae in terms of $Q_i, P_i$.

\subsection{The 2D Algebra $\left<P_1,J_1\right>$}\label{sec:Symms-P1J1}

This algebra is commutative and referred to as $\left<e_1,h_4\right>$ in \cite{f20-2,f21-1}.

It is simple to check that
\be\label{P1J1-phi}
\{\varphi,P_1\}=\{\varphi,J_1\}=0 \quad\Rightarrow\quad   \varphi = \frac{Q_2^2+Q_3^2}{k_1(Q_2^2+Q_3^2)+k_2},
\ee
with $\varphi_1(x)=k_1x+\frac{k_2}x,\;\varphi_2(y)=k_1y+\frac{k_2}y,\;\varphi_3(z)=0$.

We then consider the Hamiltonian (\ref{Hphi-QP}), with this particular $\varphi$.

\subsubsection{The Case with $\{P_1,H\}=0$}

If, in addition, we impose $\{P_1,H\}=0$, then (\ref{Hphi-QP}) reduces to
\begin{subequations}
\be\label{P1J1-H-P1}
H= \frac{Q_2^2+Q_3^2}{k_1(Q_2^2+Q_3^2)+k_2}
  \left(\frac12\left(P_1^2+P_2^2+P_3^2\right)+\frac{k_4}{Q_2^2+Q_3^2}+k_3+\frac1{Q_2^2}B\left(\frac{Q_3}{Q_2}\right)\right),
\ee
with $A_1(x)=k_3x+\frac{k_4}x,\; A_2(y)=k_3y+\frac{k_4}y$.

\smallskip
The canonical transformation, with
\be\label{P1J1canP1}
 u =Q_1,\quad v = \frac{1}{2}\, \log \left(Q_2^2+Q_3^2\right),\quad w = \arctan\left(\frac{Q_2}{Q_3}\right),
\ee
gives $P_1=p_u$ and
\be\label{P1J1-H-P1uv}
H = \frac{1}{k_1 e^{2 v}+k_2}\, \left(\frac{1}{2} (p_v^2+p_w^2) +\frac{1}{2} e^{2 v} (p_u^2+2 k_3) + k_4 + \frac{B(\cot w)}{\sin^2 w}\right).
\ee
Considered as a 2 dimensional system on the level surface $p_u=\mbox{const}$, this Hamiltonian is separable, so gives a second {\em quadratic} integral
\be\label{P1J1-K1-P1uv}
K_1 = p_w^2 +   \frac{2 B(\cot w)}{\sin^2 w}.
\ee
Any integral of (\ref{P1J1-H-P1}), which {\em commutes with} $P_1$, can be reduced to this space.  For example $K_1$ could be derived from $X_2$.
\end{subequations}

\subsubsection{The Case with $\{J_1,H\}=0$}

If, instead, we impose $\{J_1,H\}=0$, then (\ref{Hphi-QP}) reduces to
\begin{subequations}
\be\label{P1J1-H-J1}
H=\frac{Q_2^2+Q_3^2}{k_1(Q_2^2+Q_3^2)+ k_2}
  \left(\frac12\left(P_1^2+P_2^2+P_3^2\right) +\frac{A_1(x)+A_2(y)}{2R}+\frac{k_3}{Q_2^2+Q_3^2}\right),
\ee
with $B(z)=\frac{k_3}{z^2+1}$.  In this case $X_2=J_1^2+2 k_3$, so we can replace $X_2$ by $J_1$ for Liouville integrability.

\smallskip
The canonical transformation, with
\be\label{P1J1canJ1}
 u =Q_1,\quad  v = \sqrt{Q_2^2+Q_3^2},\quad   w = \arctan\left(\frac{Q_2}{Q_3}\right),
\ee
gives $J_1= -p_w$ and
\be\label{P1J1-H-J1uv}
H = \frac{v^2}{k_1 v^2 +k_2}\, \left(\frac{1}{2} (p_u^2+p_v^2) +\frac{p_w^2+2 k_3}{2 v^2} + \frac{A_1(x)+A_2(y)}{2\sqrt{u^2+v^2}}\right),
\ee
where $x=\sqrt{u^2+v^2}+u,\, y=\sqrt{u^2+v^2}-u$.

Considered as a 2 dimensional system on the level surface $p_w=\mbox{const}$, this Hamiltonian is not {\em generally} separable, but can be for particular choices of $A_i$, in which case it would again give a second {\em quadratic} integral.

Any integral of (\ref{P1J1-H-J1}), which {\em commutes with} $J_1$, can be reduced to this space.
\end{subequations}

\subsubsection{The Kepler Case (\ref{H:kepler-Q-phi}) with this $\varphi$}

It can be seen that setting $k_2=k_3=0$ and $\varphi_3\left(\frac{Q_3}{Q_2}\right)=\bar k_2$ (but then dropping the bar), that the conformal factor of (\ref{H:kepler-Q-phi}) reduces to (\ref{P1J1-phi}), giving the Hamiltonian:
\begin{subequations}
\be\label{H:kepler-Q-P1J1}
H = \frac{Q_2^2+Q_3^2}{k_1(Q_2^2+Q_3^2)+k_2} \left(\frac{1}{2}\left(P_1^2+P_2^2+P_3^2\right)-\frac{\mu}{R}+\frac1{Q_2^2}B\left(\frac{Q_3}{Q_2}\right)\right).
\ee
The remaining integrals of (\ref{HXi:kepler-Q-phi}) can similarly be reduced:
\bea
  X_1 &=& P_1S-2k_1Q_1 H-\frac{\mu Q_1}{R}, \label{X1:kepler-Q-P1J1} \\
  X_2 &=& (Q_2P_3-Q_3P_2)^2-2k_2H+\frac{2(Q_2^2+Q_3^2)}{Q_2^2}B\left(\frac{Q_3}{Q_2}\right), \label{X2:kepler-Q-P1J1} \\
  \Omega &=& -S^2+ 2k_1 R^2 H+2\mu R,  \label{Om:kepler-Q-P1J1}
\eea
\end{subequations}
If we restrict the Hamiltonian (\ref{H:kepler-Q-P1J1}) further, by demanding that $\{J_1,H\}=0$ (as in (\ref{P1J1-H-J1})), then we find that
\be\label{keplerP1J1-H}
H = \frac{Q_2^2+Q_3^2}{k_1(Q_2^2+Q_3^2)+k_2} \left(\frac{1}{2}\left(P_1^2+P_2^2+P_3^2\right)-\frac{\mu}{R}+\frac{k_3}{Q_2^2+Q_3^2}\right),
\ee
and similarly for $X_1$ and $\Omega$, since $\{J_1,X_1\}=\{J_1,\Omega\}=0$.  We find that $X_2$ can be replaced by $J_1$ itself.

If we now use the canonical transformation (\ref{P1J1canJ1}), we obtain $J_1=-p_w$ and
\begin{subequations}
\bea
   H &=& \frac{v^2}{k_1v^2+k_2}\left(\frac12(p_u^2+p_v^2)+\frac{p_w^2+2 k_3}{2v^2}-\frac{\mu}{\sqrt{u^2+v^2}}\right),  \label{keplerP1J1-Huv}  \\
   X_1 &=& p_u(up_u+vp_v)-2k_1 u H -\frac{\mu\, u}{\sqrt{u^2+v^2}},    \label{keplerP1J1-X1uv}  \\
   \Omega &=& (u p_v-vp_u)^2-\frac{2 k_2 (u^2+v^2)}{v^2}\, H + \frac{(2 k_3+p_w^2)(u^2+v^2)}{v^2}.   \label{keplerP1J1-Omuv}
\eea
\end{subequations}

\br[Kepler in $D_2$ background]
This Hamiltonian has a Darboux-Koernigs $D_2$ kinetic energy, with 2 independent quadratic integrals, but not separable in these coordinates. It is a superintegrable Kepler system in a $D_2$ geometry.  This is a deformation of the Kepler problem in 2D, with $X_1$ a Runge-Lenz integral, whilst $X_2$ is a deformation of rotational invariance.
\er

\subsubsection{The Hamiltonian (\ref{P1J1-H-J1}) with the Generalised H\'enon-Heiles Potential}

Here we consider the Hamiltonian (\ref{U2U3HX1}), which is a particular case of (\ref{P1J1-H-J1}).  The canonical transformation (\ref{P1J1canJ1}) gives
\begin{subequations}
\bea
   H &=& \frac{v^2}{k_1v^2+k_2}\left(\frac12(p_u^2+p_v^2)+\frac{p_w^2+2 k_3}{2v^2}+\frac{\omega^2}2(4u^2+v^2) +4k u(2u^2+v^2)\right),  \label{U2U3Huv}  \\
   X_1 &=& p_u(up_u+vp_v)-2k_1 u H+2\omega^2 u(2u^2+v^2)+ k (16u^4+12u^2v^2+v^4),    \label{U2U3X1uv}
\eea
\end{subequations}
in which $p_w=-J_1$, a first integral.  On each level surface $p_w=\mbox{const}$, this represents an integrable system in 2 degrees of freedom.  If $k_2=0$, this reduces to the standard ``KdV case'' of integrable H\'enon-Heiles models (see \cite{f91-1}).  When $k_1=0$, this is an extension to a constant curvature space.  For generic $k_1k_2\neq 0$, the metric is of Darboux-Koenigs type $D_2$.  When $k=0$ this potential is of ``type A'' in the classification of \cite{03-11} and the system is actually {\em super-integrable}, but the H\'enon-Heiles case, with $k\neq 0$ is just integrable.

\subsubsection{The Hamiltonian (\ref{P1J1-H-J1}) with Potential $U_4$}

If we replace $U_3$ in (\ref{U2U3HX1}) by $U_4$, we obtain another interesting subcase of (\ref{P1J1-H-J1}).

\smallskip
Specifically, if we set $A_1(x) = \frac{1}{2}\omega^2 x^3 +k x^5,\, A_2(y) = \frac{1}{2}\omega^2 y^3 +k y^5$, then
\begin{subequations}
{\small  \bea
 H &=& \frac{Q_2^2+Q_3^2}{k_1(Q_2^2+Q_3^2)+k_2}\left(\frac12\left(P_1^2+P_2^2+P_3^2\right)+\frac{\omega^2}2(4Q_1^2+Q_2^2+Q_3^2)  \right.  \nn\\
                                         &&  \hspace{5cm}     \left.     +k (16 Q_1^4+ 12 Q_1^2 (Q_2^2+Q_3^2) + (Q_2^2+Q_3^2)^2)+\frac{k_3}{Q_2^2+Q_3^2}\right), \label{U4H}\\
   X_1 &=& P_1S-2k_1Q_1H+2\omega^2Q_1(2Q_1^2+Q_2^2+Q_3^2)   \nn\\
  &&  \hspace{6cm} +  2k Q_1 \left(16Q_1^4+16Q_1^2(Q_2^2+Q_3^2)+3(Q_2^2+Q_3^2)^2\right),  \label{U4X1}
\eea    }
The canonical transformation (\ref{P1J1canJ1}) gives
\bea
   H &=& \frac{v^2}{k_1v^2+k_2}\left(\frac12(p_u^2+p_v^2)+\frac{p_w^2+2 k_3}{2v^2}+\frac{\omega^2}2(4u^2+v^2) +k (16u^4+ 12 u^2 v^2+v^4)\right),  \label{U4Huv}  \\
   X_1 &=& p_u(up_u+vp_v)-2k_1 u H+2\omega^2 u(2u^2+v^2)+ 2 k u (16u^4+16u^2v^2+3 v^4),    \label{U4X1uv}
\eea
in which $p_w=-J_1$, a first integral.

Again, as a 2 dimensional sytem and for generic $k_1k_2\neq 0$, the metric is of Darboux-Koenigs type $D_2$.  When $k=0$ it is a ``type A'' potential in the classification of \cite{03-11}.  The only difference between (\ref{U2U3Huv}) and (\ref{U4Huv}) is that the H\'enon-Heiles part of the potential has been replaced by the quartic potential in the 2 dimensional ``parabolic'' series.

\end{subequations}

\subsection{The 2D Algebra $\left<S,J_1\right>$}\label{sec:Symms-SJ1}

This algebra is commutative and would be referred to as $\left<h_1,h_4\right>$ in \cite{f20-2,f21-1}, which, in that context, is equivalent to $\left<h_1,h_2\right>$ in our classification.  However, it is clear from the formulae (\ref{HX1X2phi-QP}) that $J_1$ ($=h_4$) acts in a special way, so this equivalence no longer holds.

Since $S$ acts {\em conformally} on $H^0 = \frac{1}{2} (P_1^2+P_2^2+P_3^2)$, we need to consider $\varphi H^0$ instead of $\varphi$:

\be\label{SJ1-phi}
\{\varphi H^0,S\}=\{\varphi H^0,J_1\}=0 \quad\Rightarrow\quad   \varphi = \frac{(Q_2^2+Q_3^2) R}{k_1 R+k_2 Q_1},
\ee
where $k_1=\frac{1}{2} (\bar k_1+\bar k_2+2\bar k_3),\; k_2=\frac{1}{2} (\bar k_2-\bar k_1)$ and $\varphi_1= \frac{\bar k_1}{x},\;\varphi_2= \frac{\bar k_2}{y},\;\varphi_3(z)=\bar k_3$.

\subsubsection{The Case with $\{S,H\}=0$}

If, in addition, we impose $\{S,H\}=0$, then (\ref{Hphi-QP}) reduces to
\begin{subequations}
\be\label{SJ1-H-S}
H= \frac{(Q_2^2+Q_3^2) R}{k_1 R+k_2 Q_1}
  \left(\frac12\left(P_1^2+P_2^2+P_3^2\right)+\frac{k_3+k_4}{2(Q_2^2+Q_3^2)}+\frac{(k_4-k_3)Q_1}{2(Q_2^2+Q_3^2)R}+\frac1{Q_2^2}B\left(\frac{Q_3}{Q_2}\right)\right),
\ee
with $A_1(x)=\frac{k_3}x,\; A_2(y)=\frac{k_4}y$.

\smallskip
The canonical transformation, with
\be\label{SJ1canS}
 u = \frac{1}{2}\, \log \left(Q_1^2+Q_2^2+Q_3^2\right),\quad v = \log \left(\frac{Q_1+R}{\sqrt{Q_2^2+Q_3^2}}\right),\quad w = \arctan\left(\frac{Q_2}{Q_3}\right),
\ee
gives $S=p_u$ and
\be\label{SJ1-H-Suv}
H = \frac{1}{k_1 +k_2 \tanh v}\, \left(\frac{1}{2} (p_v^2+p_w^2) +\frac{p_u^2}{2 \cosh^2 v} + \frac{1}{2} (k_3+ k_4) + \frac{1}{2} (k_4- k_3) \tanh v  + \frac{B(\cot w)}{\sin^2 w}\right).
\ee
Considered as a 2 dimensional system on the level surface $p_u=\mbox{const}$, this Hamiltonian is separable, so gives a second {\em quadratic} integral
\be\label{SJ1-K1-Suv}
K_1 = p_w^2 +   \frac{2 B(\cot w)}{\sin^2 w}.
\ee
Any integral of (\ref{SJ1-H-S}), which {\em commutes with} $S$, can be reduced to this space.  For example $K_1$ could be derived from $X_2$.
\end{subequations}

\subsubsection{The Case with $\{J_1,H\}=0$}

If, instead, we impose $\{J_1,H\}=0$, then (\ref{Hphi-QP}) reduces to
\begin{subequations}
\be\label{SJ1-H-J1}
H= \frac{(Q_2^2+Q_3^2) R}{k_1 R+k_2 Q_1}
  \left(\frac12\left(P_1^2+P_2^2+P_3^2\right) +\frac{A_1(x)+A_2(y)}{2R}+\frac{k_3}{Q_2^2+Q_3^2}\right),
\ee
with $B(z)=\frac{k_3}{z^2+1}$.

\smallskip
The canonical transformation, with
\be\label{SJ1canJ1}
 u = \frac{1}{2}\, \log \left(Q_1^2+Q_2^2+Q_3^2\right),\quad v =  \arctan \left(\frac{Q_1}{\sqrt{Q_2^2+Q_3^2}}\right),\quad w = \arctan\left(\frac{Q_2}{Q_3}\right),
\ee
gives $J_1=-p_w$ and
\be\label{SJ1-H-J1uv}
H = \frac{\cos^2 v}{k_1 +k_2 \sin v}\, \left(\frac{1}{2} (p_u^2+p_v^2) +\frac{p_w^2+2 k_3}{2 \cos^2 v} + \frac{1}{2} e^u (A_1(x)+A_2(y))\right),
\ee
where $x=e^u (1+\sin v),\, y=e^u (1-\sin v)$

Considered as a 2 dimensional system on the level surface $p_w=\mbox{const}$, this Hamiltonian is not {\em generally} separable, but can be for particular choices of $A_i$, in which case it would again give a second {\em quadratic} integral.

Any integral of (\ref{SJ1-H-J1}), which {\em commutes with} $J_1$, can be reduced to this space.
\end{subequations}

\subsubsection{The Kepler Case (\ref{H:kepler-Q-phi}) with this $\varphi$}

It can be seen that setting $k_1=k_3=0$ and $\varphi_3\left(\frac{Q_3}{Q_2}\right)=\bar k_1,\, k_2=\bar k_2$ in the conformal factor (\ref{phi:kepler-Q-phi})  (but dropping the bars), then (\ref{H:kepler-Q-phi}) reduces to (\ref{SJ1-phi}), giving the Hamiltonian:
\begin{subequations}
\be\label{H:kepler-Q-SJ1}
H =  \frac{(Q_2^2+Q_3^2) R}{k_1 R+k_2 Q_1} \left(\frac{1}{2}\left(P_1^2+P_2^2+P_3^2\right)-\frac{\mu}{R}+\frac1{Q_2^2}B\left(\frac{Q_3}{Q_2}\right)\right).
\ee
The remaining integrals of (\ref{HXi:kepler-Q-phi}) can similarly be reduced:
\bea
  X_1 &=& P_1S+k_2\, \frac{H}{R}-\frac{\mu Q_1}{R}, \label{X1:kepler-Q-SJ1} \\
  X_2 &=& (Q_2P_3-Q_3P_2)^2 +\frac{2(Q_2^2+Q_3^2)}{Q_2^2}B\left(\frac{Q_3}{Q_2}\right), \label{X2:kepler-Q-SJ1} \\
  \Omega &=& -S^2 +2\mu R,  \label{Om:kepler-Q-SJ1}
\eea
\end{subequations}
If we restrict the Hamiltonian (\ref{H:kepler-Q-SJ1}) further, by demanding that $\{J_1,H\}=0$ (as in (\ref{SJ1-H-J1})), then we find that
\be\label{keplerSJ1-H}
H = \frac{(Q_2^2+Q_3^2) R}{k_1 R+k_2 Q_1} \left(\frac{1}{2}\left(P_1^2+P_2^2+P_3^2\right)-\frac{\mu}{R}+\frac{k_3}{Q_2^2+Q_3^2}\right),
\ee
and similarly for $X_1$ and $\Omega$.  We find that $X_2$ can be replaced by $J_1$ itself.

The canonical transformation (\ref{SJ1canJ1}) gives the following system in 2 dimensions (for $p_w$ a constant)
\begin{subequations}
\bea
H &=& \frac{\cos^2 v}{k_1 +k_2 \sin v}\,\left(\frac{1}{2} (p_u^2+p_v^2)+\frac{p_w^2+2 k_3}{2\cos^2 v} - \mu e^u\right),   \label{H:kepler-uv-SJ1}  \\
X_1 &=&  e^{-u}\, p_u \left(p_u \sin v + p_v \cos v\right) + k_2 e^{-u}\, H -\mu \sin v,  \label{X1:kepler-uv-SJ1}  \\
\Omega &=&  -(p_u^2-2\mu\, e^{u}).
\eea
In these coordinate, the Hamiltonian is separable, so gives a second quadratic integral, which is just $- \Omega$.

For generic $k_i$, the Hamiltonian (\ref{H:kepler-uv-SJ1}) has a Darboux-Koenigs type $D_4$ kinetic energy, with a potential of ``type A'' in the classification of \cite{03-11}.  It reduces to a constant curvature space, when $k_2=0$.
\end{subequations}

\br[Other Potentials]
In the $Q_i$ coordinates, {\em any} of the potentials considered in Section \ref{sec:Symms-P1J1} can be written in the case of (\ref{SJ1-H-J1}).  However, the canonical transformation (\ref{SJ1canJ1}) is based on the symmetry algebra of the \underline{kinetic energy}, so only special potentials will be well adapted to these.  In \cite{f21-1} we concentrated on {\em separable} potentials, but as seen in (\ref{keplerP1J1-Huv}), there are other potentials which give simple and interesting forms.
\er

\subsection{A Copy of the 3D Algebra $\mathfrak{sl}(2)$}\label{sec:Symms-sl2}

Here we consider a particular realisation of $\mathfrak{sl}(2)$, within the conformal algebra, which (in \cite{f20-2,f21-1}) we label $\left<e_1,h_1,f_1\right>$:
\be\label{sl2}
\left. \begin{array}{l}
e_1=P_1,\;\; h_1 = S,\\[2mm]
 f_1 = (Q_1^2+Q_2^2+Q_3^2) P_1 -2 Q_1 S
   \end{array} \right\}  \quad\Rightarrow\quad \{h_1,e_1\}=e_1,\;\; \{h_1,f_1\}= -f_1,\;\; \{e_1,f_1\}= 2 h_1.
\ee

It is simple to check that
\be\label{sl2-phi}
\{\varphi H^0,P_1\}=\{\varphi H^0,S\}=0 \quad\Rightarrow\quad   \varphi = \frac{Q_2^2+Q_3^2}{\varphi_3\left(\frac{Q_3}{Q_2}\right)+k_1},
\ee
where $\varphi_1= \frac{k_1}{x},\;\varphi_2= \frac{k_1}{y}$.  It is then simple to check that $\{\varphi H^0,f_1\}=0$, without further constraint.

\subsubsection{The Case with $\{P_1,H\}=0$}

If, in addition, we impose $\{P_1,H\}=0$, then (\ref{Hphi-QP}) reduces to
\begin{subequations}
\be\label{sl2-H-P1}
H= \frac{(Q_2^2+Q_3^2)}{\varphi_3\left(\frac{Q_3}{Q_2}\right)+k_1}
  \left(\frac12\left(P_1^2+P_2^2+P_3^2\right)+\frac{k_3}{Q_2^2+Q_3^2}+k_2+\frac1{Q_2^2}B\left(\frac{Q_3}{Q_2}\right)\right),
\ee
with $A_1(x)=k_2x+\frac{k_3}x,\; A_2(y)=k_2y+\frac{k_3}y$.

\smallskip
The canonical transformation, with
\be\label{sl2canP1}
 u = Q_1, \quad v = \arctan\left(\frac{Q_3}{Q_2}\right),\quad w = \frac{1}{2}\, \log (Q_2^2+Q_3^2),
\ee
gives $P_1=p_u$ and
\be\label{sl2-H-P1uv}
H = \frac{1}{k_1 +\varphi_3 (\tan v)}\, \left(\frac{1}{2} (p_v^2+p_w^2) +\frac{1}{2}\, e^{2 w} (p_u^2+2 k_2) + k_3 +  \frac{B(\tan v)}{\cos^2 v}\right).
\ee
Considered as a 2 dimensional system on the level surface $p_u=\mbox{const}$, this Hamiltonian is separable, so gives a second {\em quadratic} integral
\be\label{sl2-K1-P1uv}
K_1 = p_w^2 + e^{2 w} (p_u^2+2 k_2).
\ee
\end{subequations}
Any integral of (\ref{sl2-H-P1}), which {\em commutes with} $P_1$, can be reduced to this space.  For example $K_1$ could be derived from $2 k_1 H-X_2-2 k_3$.

\subsubsection{The Case with $\{S,H\}=0$}

If, instead, we impose $\{S,H\}=0$, then (\ref{Hphi-QP}) reduces to
\begin{subequations}
\be\label{sl2-H-S}
H= \frac{(Q_2^2+Q_3^2)}{\varphi_3\left(\frac{Q_3}{Q_2}\right)+k_1}
  \left(\frac12\left(P_1^2+P_2^2+P_3^2\right)+\frac{k_3+k_2}{2(Q_2^2+Q_3^2)}+\frac{(k_3-k_2)Q_1}{2(Q_2^2+Q_3^2)R}+\frac1{Q_2^2}B\left(\frac{Q_3}{Q_2}\right)\right),
\ee
with $A_1(x)=\frac{k_2}x,\; A_2(y)=\frac{k_3}y$.

\smallskip
The canonical transformation, with
\be\label{sl2canS}
 u = \frac{1}{2}\, \log \left(\frac{R+Q_1}{R-Q_1}\right), \quad v = \arctan\left(\frac{Q_3}{Q_2}\right), \quad w =  \frac{1}{2}\, \log (Q_1^2+Q_2^2+Q_3^2),
\ee
gives $S=p_w$ and
\be\label{sl2-H-Suv}
H = \frac{1}{k_1 +\varphi_3 (\tan v)}\, \left(\frac{1}{2} (p_u^2+p_v^2) +\frac{p_w^2}{2 \cosh^2 u} +\frac{1}{2}(k_3+k_2) +\frac{1}{2} (k_3-k_2) \tanh u  +  \frac{B(\tan v)}{\cos^2 v}\right).
\ee
Considered as a 2 dimensional system on the level surface $p_w=\mbox{const}$, this Hamiltonian is separable, so gives a second {\em quadratic} integral
\be\label{sl2-K1-Suv}
K_1 = p_u^2 +\frac{p_w^2}{\cosh^2 u} + (k_3-k_2) \tanh u.
\ee
\end{subequations}
Any integral of (\ref{sl2-H-S}), which {\em commutes with} $S$, can be reduced to this space.  For example $K_1$ could be derived from $2 k_1 H-X_2-k_2- k_3$.

\br[Lie algebra involution]
The case of $\{f_1,H\}=0$ {\em looks} more complicated, but is related to the $\{P_1,H\}=0$ case through an involution of $\mathfrak{sl}(2)$, realised by
$$
(Q_1,Q_2,Q_3) \mapsto \left(\frac{Q_1}{R^2},\frac{Q_2}{R^2},\frac{Q_3}{R^2}\right) \quad\Rightarrow\quad e_1 \leftrightarrow f_1, \;\; h_1 \mapsto -h_1.
$$
\er

\subsection{The 3D Algebra $\mathfrak{so}(3)$}\label{sec:Symms-so3}

Here we consider the rotational symmetries of $H^0 = \frac{1}{2} (P_1^2+P_2^2+P_3^2)$.  In \cite{f20-2,f21-1} this algebra is called $\left<h_2,h_3,h_4\right>$.

It is simple to check that
\be\label{phi-so3}
\{\varphi,J_1\}=\{\varphi,J_2\}=\{\varphi,J_3\}=0 \quad\Rightarrow\quad   \varphi = \frac{R}{k_1 R+k_2},
\ee
with $\varphi_1= k_1 x + k_2,\;\varphi_2= k_1 y + k_2,\;\varphi_3= 0$.

\subsubsection{The Case with $\{J_1,H\}=0$}

If, in addition, we impose $\{J_1,H\}=0$, then (\ref{Hphi-QP}) reduces to
\begin{subequations}
\be\label{so3-H-J1}
H= \frac{R}{k_1 R+k_2} \left(\frac12\left(P_1^2+P_2^2+P_3^2\right) +\frac{A_1(x)+A_2(y)}{2R}+\frac{k_3}{Q_2^2+Q_3^2}\right),
\ee
with $B(z)=\frac{k_3}{z^2+1}$.  With these values of $\varphi_i$ and $A_i$, the functions $H, X_1, X_2$ are still in involution.  However, since $X_2=J_1^2+2k_3$, we can just replace it by $J_1$.

\smallskip
The canonical transformation, with
\be\label{so3canJ1}
u = \arctan\left(\frac{\sqrt{Q_2^2+Q_3^2}}{Q_1}\right),\quad v = \frac{1}{2}\, \log \left(Q_1^2+Q_2^2+Q_3^2\right),\quad w =  \arctan \left(\frac{Q_2}{Q_3}\right),
\ee
adapted to $J_1$, with $p_w=-J_1$, gives the Hamiltonian
\be\label{so3-H-J1uv}
H =  \frac{e^{-v}}{k_1 e^v+k_2} \left( \frac{1}{2}\, (p_u^2+p_v^2) +\frac{p_w^2+2 k_3}{2 \sin^2 u} + \frac{1}{2}\,e^v (A_1(x)+A_2(y))\right),
\ee
\end{subequations}
where $x=e^v (1+\cos u),\, y=e^v (1-\cos u)$.

Considered as a 2 dimensional system on the level surface $p_w=\mbox{const}$, this Hamiltonian is not {\em generally} separable, but can be for particular choices of $A_i$, in which case it would again give a second {\em quadratic} integral.

Any integral of (\ref{so3-H-J1}), which {\em commutes with} $J_1$, can be reduced to this space.

\subsubsection{The Case with $\{J_2,H\}=0$}

If, instead, we impose $\{J_2,H\}=0$, then (\ref{Hphi-QP}) reduces to
\begin{subequations}
\be\label{so3-H-J2}
H= \frac{R}{k_1 R+k_2}
  \left(\frac12\left(P_1^2+P_2^2+P_3^2\right) +k_3+\frac{k_4}{2R}+\frac{k_5}{Q_2^2}\right),
\ee
with $A_1(x)=k_3 x,\quad A_2(y)=k_3 y+k_4,\quad B(z)= k_5$.

\smallskip
The canonical transformation, with
\be\label{so3canJ2}
u = \arctan\left(\frac{\sqrt{Q_1^2+Q_3^2}}{Q_2}\right),\quad v = \frac{1}{2}\, \log \left(Q_1^2+Q_2^2+Q_3^2\right),\quad w =  \arctan \left(\frac{Q_3}{Q_1}\right),
\ee
adapted to $J_2$, with $p_w=- J_2$, gives a separable form of (\ref{so3-H-J2}):
\be\label{so3-H-J2uv}
H = \frac{e^{-v}}{k_1 e^v+k_2} \left( \frac{1}{2}\, (p_u^2+p_v^2) +\frac{p_w^2}{2\sin^2 u} + k_3 e^{2v}+\frac{1}{2}\, k_4 e^v + k_5 \sec^2 u\right),
\ee
which is equivalent (on the level surface $p_w=\mbox{const}$) to the Darboux-Koenigs $D_3$ kinetic energy, with a potential.  In this separable form, it can be seen that there is a {\em quadratic} integral
\be\label{so3-K1-J2uv}
K_1 = p_u^2 +\frac{p_w^2}{\sin^2 u} + 2 k_5 \sec^2 u.
\ee
Any integral of (\ref{so3-H-J2}) that \underline{commutes with} $J_2$ can be reduced to this space. In particular, since $p_u^2 +\frac{p_w^2}{\sin^2 u}=J^2$, $K_1$ is related to a deformation of the rotational Casimir, which is an additional integral in this case.
\end{subequations}

\br[The Case with $\{J_3,H\}=0$]

Similarly, if we impose $\{J_3,H\}=0$, then (\ref{Hphi-QP}) reduces to
$$
H= \frac{R}{k_1 R+k_2}
  \left(\frac12\left(P_1^2+P_2^2+P_3^2\right) +k_3+\frac{k_4}{2R}+\frac{k_5}{Q_3^2}\right),
$$
with $A_1(x)=k_3 x,\quad A_2(y)=k_3 y+k_4,\quad B(z)= \frac{k_5}{z^2}$.
\er

\subsubsection{Separable Case of (\ref{so3-H-J1uv})}\label{sec:so3-J1-sep}

For general $A_1(x)$ and $A_2(y)$, there will be no additional first integrals of (\ref{so3-H-J1}), but special choices can render it {\em superintegrable}.

First, we consider separability of (\ref{so3-H-J1uv}), which requires
\begin{subequations}
\be\label{so3-H-J1-sep}
A_1(x)+A_2(y) = a(v)+e^{-v} b(u),
\ee
for some functions $a(v),\, b(u)$.  If we differentiate (\ref{so3-H-J1-sep}) with respect to $u$ and also with respect to $v$, we obtain simultaneous equations for $A_i'$, giving (for $A_1(x)$)
\be\label{so3-H-J1-A1x}
2 e^{2 v} A_1'(x) =e^v a'(v)+b(u) +\tan\left(\frac{u}{2}\right)\, b'(u) \quad\Rightarrow\quad   \left(2 e^{2 v} A_1'(x)\right)_{uv} = 0.
\ee
This, together with a similar equation for $A_2(y)$, give
\be\label{so3-H-J1-A1A2eq}
x A_{1xxx}+3 A_{1xx}=0,\;\;\; y A_{2yyy}+3 A_{2yy} = 0 \quad\Rightarrow\quad A_1=a_0+a_1 x+\frac{a_2}{x},\;\; A_2=b_0+b_1 y+\frac{b_2}{y}.
\ee
Substituting these back into (\ref{so3-H-J1-sep}) requires one condition ($b_1=a_1$) on these 6 parameters, giving
\be\label{A1+A2sol}
\frac{1}{2} \, e^v (A_1(x)+A_2(y)) = \frac{1}{2} (a_0+b_0) e^v +a_1 e^{2 v} +\frac{a_2}{2(1+\cos u)}  +\frac{b_2}{2(1-\cos u)}.
\ee
\end{subequations}
The $e^{2 v}$ term can be absorbed into the $e^v$ term by noting that
$$
a_1 e^{2 v} = \frac{a_1}{k_1} \, k_1 e^{2 v} \quad\mbox{and}\quad  \left(\frac{e^{-v}}{k_1 e^v+k_2} \right) \left(\frac{a_1}{k_1} \left(k_1 e^{2 v}+k_2 e^v\right)\right) = \frac{a_1}{k_1},
$$
so, without loss of generality, we can set $a_1=0$ and
\begin{subequations}
\be\label{A1+A2sol2}
\frac{1}{2} \, e^v (A_1(x)+A_2(y)) = \frac{1}{2} (a_0+b_0) e^v  +\frac{1}{2} (b_2-a_2)\,\frac{\cos u}{\sin^2 u}  +\frac{a_2+b_2}{2\sin^2u}.
\ee
The last term can be absorbed into the $\frac{k_3}{\sin^2 u}$ term of (\ref{so3-H-J1uv}), so, defining $\beta$ and $\gamma$ by
$$
a_0+b_0  = 2 \beta,  \quad  b_2 -a_2 = 2 \gamma,
$$
this leads to
\be\label{so3-Hsep-J1}
H =  \frac{e^{-v}}{k_1 e^v+k_2} \left( \frac{1}{2}\, (p_u^2+p_v^2) +\frac{p_w^2+2 k_3}{2 \sin^2 u} +\beta e^v + \gamma\, \frac{\cos u}{\sin^2 u}\right).
\ee

From separability, we see that
\be\label{so3-H-J1-K1uv}
K_1 = p_u^2 +\frac{p_w^2+2 k_3}{\sin^2 u} + 2\gamma\, \frac{\cos u}{\sin^2 u}
\ee
is a first integral.  Since $p_u^2 +\frac{p_w^2}{\sin^2 u}$ corresponds to the rotational Casimir $J^2$, $K_1$ is an additional integral.

\smallskip
We also have that $\{H,X_1\}=0$ (in $Q_1$ coordinates), so $X_1$ can be reduced to $u-v$ coordinates:
\be\label{so3-H-J1-X1uv}
X_1 = e^{-v} p_v \left(p_v \cos u-p_u \sin u\right) -(2 k_1 e^v +k_2) \cos u\, H + \beta \cos u - \gamma e^{-v}.
\ee
The Hamiltonian (\ref{so3-Hsep-J1}) is the Darboux-Koenigs $D_3$ kinetic energy, with potential of `type B' in the classification of \cite{03-11}, so not just separable, but \underline{maximally superintegrable}.

\end{subequations}

\subsubsection{Superintegrability of (\ref{so3-H-J2})}\label{sec:so3-H-J2-super}

With the choice of $\varphi_i, A_i$ and $B$ for this case, we find (removing an additive constant from $X_1$)
\begin{subequations}
\bea
X_1 &=& P_1 S - \frac{Q_1 (k_2+2 k_1 R)}{R}\, H + Q_1 \left( 2 k_3+\frac{k_4}{2R}\right),  \label{so3-H-J2-X1}  \\
X_2 &=& J_1^2+\frac{2 k_5 (Q_2^2+Q_3^2)}{Q_2^2} ,  \label{so3-H-J2-X2}
\eea
with $\{X_1,X_2\}=0$ and
\bea
X_3 &=& \{X_1,J_2\} = P_3 S- Q_3 \left(\frac{k_2+2 k_1 R}{R}\right)\, H+ Q_3 \left( 2 k_3+\frac{k_4}{2R}\right), \label{so3-H-J2-X3}  \\
X_4 &=& \frac{1}{2}\, \{X_2,J_2\} = J_1 J_3 - 2 k_5 \frac{Q_1 Q_3}{Q_2^2}.   \label{so3-H-J2-X4}
\eea
\end{subequations}
It is easy to check that $H, X_1, X_2, J_2$, together with either of the quantities $\{X_i,J_2\}$, have rank 5, so this is a maximally superintegrable system.  It is a conformally flat extension of the second system in Table I of \cite{90-22}.

Whilst we do not present the full Poisson algebra, there are some further integrals, which are useful below:
\begin{subequations}
\bea
 X_5 &=& \{X_4,J_2\}= J_3^2-J_1^2+ 2 k_5 \left( \frac{Q_1^2-Q_3^2}{Q_2^2}\right),    \label{so3-H-J2-X5}  \\
 X_6 &=& J^2 + 2 k_6 \left( \frac{Q_1^2+Q_3^2}{Q_2^2}\right),  \label{so3-H-J2-X6}
\eea
with $\{X_6,J_2\}=0$.  These also satisfy
\bea
 \{X_3,J_2\}=-X_1 & \Rightarrow & \{F_1,J_2\}=0,\;\;\; \mbox{where}\;\;\; F_1=X_1^2+X_3^2, \label{so3-H-J2-F1} \\
  \{X_5,J_2\}=-4X_4 & \Rightarrow & \{F_2,J_2\}=0,\;\;\; \mbox{where}\;\;\; F_2= 4X_4^2+X_5^2,     \label{so3-H-J2-F2}
\eea
so $F_1$ and $F_2$ are {\em quartic} integrals which commute with $J_2$.
\end{subequations}

\subsubsection*{2D Reduction}

The canonical transformation (\ref{so3canJ2}) transforms (\ref{so3-H-J2}) into the form (\ref{so3-H-J2uv}), with quadratic integral $K_1$, given by (\ref{so3-K1-J2uv}).
As previously said, any integral of (\ref{so3-H-J2}) that \underline{commutes with} $J_2$ can be reduced to this space.  In fact $X_6 = K_1-2 k_5$.

Whilst this potential is very {\em similar} to `type B' in the classification of \cite{03-11}, it has significant differences and it can be shown that there are no further \underline{quadratic} integrals of (\ref{so3-H-J2uv}).  However, we have the two {\em quartic} integrals $F_1$ and $F_2$.  In fact, $F_2$ is related to $K_1$, in that $F_2=(K_1-p_w^2)^2-4 k_5 (K_1+p_w^2-k_5)$, where $p_w$ is just a parameter in this context.  However, $H, K_1, F_1$ are functionally independent, so this 2D system is maximally superintegrable, but \underline{not} in the class studied in \cite{03-11}.

\subsubsection{The Kepler Case (\ref{H:kepler-Q-phi}) with this $\varphi$}\label{sec:Kepler-so3}

It can be seen that setting $k_2=0,\, \varphi_3\left(\frac{Q_3}{Q_2}\right)=0$ and $k_3= 2 \bar k_2$ in the conformal factor (\ref{phi:kepler-Q-phi})  (but dropping the bar), then (\ref{phi:kepler-Q-phi}) reduces to (\ref{phi-so3}), giving the Hamiltonian:
\begin{subequations}
\be\label{so3-H-kepler}
H =  \frac{R}{k_1 R+k_2} \left(\frac{1}{2}\left(P_1^2+P_2^2+P_3^2\right)-\frac{\mu}{R}+\frac1{Q_2^2}B\left(\frac{Q_3}{Q_2}\right)\right).
\ee
The remaining integrals of (\ref{HXi:kepler-Q-phi}) can similarly be reduced:
\bea
  X_1 &=& P_1S- \frac{Q_1 (2 k_1 R+k_2)}{R} \, H -\frac{\mu Q_1}{R}, \label{so3-X1-kepler} \\
  X_2 &=& (Q_2P_3-Q_3P_2)^2 +\frac{2(Q_2^2+Q_3^2)}{Q_2^2}B\left(\frac{Q_3}{Q_2}\right), \label{so3-X2-kepler} \\
  \Omega &=& -S^2 +2 (k_1 R^2+k_2 R) H +2\mu R.  \label{so3-Om-kepler}
\eea
\end{subequations}

\paragraph{For $\{J_1,H\}=0$} we have $B\left(\frac{Q_3}{Q_2}\right)=\frac{k_3 Q_2^2}{Q_2^2+Q_3^2}$, reducing (\ref{so3-H-kepler}) to a particular case of (\ref{so3-H-J1}), which reduces to a particular example of the separable case, (\ref{so3-Hsep-J1}), with $a_2=\gamma=0$:
\be\label{so3-H-kepler-J1uv}
H =  \frac{e^{-v}}{k_1 e^v+k_2} \left( \frac{1}{2}\, (p_u^2+p_v^2) +\frac{p_w^2+2 k_3}{2 \sin^2 u} -\mu e^v\right) ,
\ee
with corresponding particular case of $K_1$ and $X_1$.  The function $K_1$ just corresponds to $\Omega$ in these coordinates.

\paragraph{For $\{J_2,H\}=0$} we have $B\left(\frac{Q_3}{Q_2}\right)=k_3$, reducing (\ref{so3-H-kepler}) to a particular case of (\ref{so3-H-J2}), and hence reduces to a particular case of (\ref{so3-H-J2uv}):
\be\label{so3-H-kepler-J2uv}
H = \frac{e^{-v}}{k_1 e^v+k_2} \left( \frac{1}{2}\, (p_u^2+p_v^2) +\frac{p_w^2}{2\sin^2 u} -\mu e^v + k_3 \sec^2 u\right),
\ee
with corresponding particular case of $K_1$, which corresponds to $\Omega$ in these coordinates.  As we found in Section \ref{sec:so3-H-J2-super}, $X_1$ does not reduce to these coordinates and, indeed, that there are \underline{no} further {\em quadratic} integrals.

However, we do have the {\em quartic} integral $F_1$ and, as previously remarked, $H, K_1, F_1$ are functionally independent, so this 2D system is maximally superintegrable, but \underline{not} in the class studied in \cite{03-11}.

\section{A Conformally Flat  Kepler-Calogero-Moser System}\label{sec:conf-KepCM}

In Section \ref{sec:Kepler-Q-phi} we presented the general conformally flat version of the Kepler case in $(Q_i,P_i)$ coordinates.  This involved the arbitrary function $B\left(\frac{Q_3}{Q_2}\right)$, which, in the particular Calogero-Moser case, should take the form $B(z) = \frac{9 g^2 (1+z^2)^2}{2 (1-3 z^2)^2}$, as already discussed in the flat case of Section \ref{sec:KCM}.  Here we just present the case with a spherically symmetric kinetic energy, given in Section \ref{sec:Kepler-so3}:
\begin{subequations}
\bea
  H &=& \frac{r}{k_1r+k_2}\left(\frac12\left(p_1^2+p_2^2+p_3^2\right)+g^2\left(\frac1{(q_1-q_2)^2}+\frac1{(q_1-q_3)^2}+\frac1{(q_2-q_3)^2}\right) -\frac{\mu}{r}\right), \label{cflat-H-CMK} \\
  X_1 &=& (p_1+p_2+p_3)S-\frac{1}{r} \left((q_1+q_2+q_3)(2k_1r+k_2)\, H+\mu(q_1+q_2+q_3)\right),   \label{cflat-X1-CMK}  \\
  X_2 &=& (J_1+J_2+J_3)^2    \nn  \\
    &&  +4g^2(q_1^2+q_2^2+q_3^2-q_1q_2-q_2q_3-q_1q_3)\left(\frac1{(q_1-q_2)^2}+\frac1{(q_1-q_3)^2}+\frac1{(q_2-q_3)^2}\right),   \label{cflat-X2-CMK}  \\
  \Omega &=& -S^2+2r\left(k_1 r+k_2\right)H+2\mu\, r,    \label{cflat-Om-CMK}
\eea
where $r=\sqrt{q_1^2+q_2^2+q_3^2}$.  As before, we need to add a cubic integral, $\{X_1,\Omega\}=-2X_3$, to this list, given by:
\bea
 X_3 &=& (p_1+p_2+p_3) \Omega    -\frac{\mu+k_2 H}{\sqrt{q_1^2+q_2^2+q_3^2}} \, ((q_2^2+q_3^2-q_1(q_2+q_3))p_1  \nn\\
     &&  \qquad  +(q_3^2+q_1^2-q_2(q_3+q_1))p_2+(q_1^2+q_2^2-q_3(q_1+q_2))p_3).  \label{cflat-X3-CMK}
\eea
\end{subequations}

These functions satisfy the Poisson relations:
\begin{subequations}
\be
  \{X_1,X_3\}=(2k_1 (X_2-6\Omega)-3k_2^2H)H+X_1^2-6\mu k_2 H-3\mu^2,\qquad \{\Omega,X_3\}=-2X_1\Omega,  \label{cflat-XiXj}
\ee
together with the constraint
\be
  X_3^2=(3k_2^2H^2+2k_1(3\Omega-X_2)H-X_1^2+6\mu k_2H+3\mu^2)\Omega-(k_2H+\mu)^2X_2.   \label{cflat-X3^2=}
\ee
\end{subequations}
In the reduction $k_1=1,\, k_2=0$, these integrals and relations just reduce to those of Section \ref{sec:KCM}.

\section{Conclusions}

This paper has continued the work of \cite{f20-2,f21-1}.  Here we derive the consequences of having one particular {\em quadratic} first integral $X_1$, which allows the Calogero-Moser model as a particular case.  This led to the Liouville integrable triple (\ref{HX1X2Q}), in terms of three arbitrary functions $A_1, A_2$ and $B$.  The functions $A_i$ correspond to a 3D generalisation of one of the standard 2D separable systems (in parabolic coordinates).  The function $B$ allows for the connection to the Calogero-Moser system, corresponding to a particular choice of $B$.  This gives a coupling of the Calogero-Moser system with a potential depending on \underline{2 arbitrary functions}, which includes (for special choices of $A_i$) the Kepler potential, a resonant harmonic oscillator, the (KdV related) H\'enon-Heiles potential and many more.  All of these are \underline{at least} Liouville integrable, but, in many cases, \underline{superintegrable}.

\smallskip
This whole class of system was extended to the ``conformally flat'' case, including the Kepler-Calogero-Moser system with a spherically symmetric kinetic energy.  However, the most interesting results came through symmetry reduction, which mostly destroyed the connection with the Calogero-Moser system. For example we presented a Kepler problem (\ref{keplerP1J1-Huv}) and a generalised H\'enon-Heiles system (\ref{U2U3Huv}) on a Darboux-Koenigs $D_2$ background.  Other systems reduced to a $D_1,\, D_3$ or $D_4$ backgrounds.

Two and three dimensional systems have been studied a lot, so are quite well understood.  Some particular cases are known for \underline{general} $n$, but these are rare, mainly associated with Lax pairs (such as the rational Calogero-Moser system and its generalisations).  However, it would be interesting to investigate how the construction of this paper could be generalised to 4 and higher dimensions.  In particular, the conformally flat case should lead to interesting new features.  As emphasised in \cite{f19-3,f19-2,f20-2}, the conformal algebra plays an important role in building both Liouville and superintegrable systems.  It is particularly important in the non-constant curvature case, where Killing tensors (the leading order coefficients in higher order first integrals) cannot always be built from Killing vectors.  Symmetry reductions (as in Section \ref{sec:Symms-phi}) should be particularly interesting, allowing us to reduce from 4 to 3 to 2 dimensions in a variety of ways.

Of course, the hope is that those systems which exist for {\em all} $n$, will have a Lax pair.  However, constructing this Lax pair could be a difficult task.

Most superintegrable systems in classical mechanics have quantum counterparts, although some problems can occur with higher order integrals \cite{05-9,f19-3}.  Most of the emphasis is on deriving the spectrum and solving for eigenfunctions through separation of variables \cite{71-3,96-5,17-5}.  However, as emphasised in \cite{f07-1,f20-3}, superintegrability can be \underline{directly} used for building eigenfunctions.  For many of the systems discussed in this paper, the analysis of the quantum version is an open problem.

\subsection*{Acknowledgements}

This work was supported by the National Natural Science Foundation of China (grant no. 11871396).

\section{Appendix}\label{sec:appendix}

\subsection{Notation and Definitions}\label{sec:appendix-not}

Here we collect some useful definitions and formulae, used throughout the paper.

We give a detailed analysis of the conformal algebra in 3D in \cite{f20-2}, but here just emphasise a few important elements.  For our purposes it is more convenient to write all formulae in terms of Hamiltonian functions, rather than as metric tensors and Killing vectors.

The symmetry algebra (a subalgebra of the conformal algebra) of the Kinetic energy $H^0 = \frac{1}{2} (p_1^2+p_2^2+p_3^2)$ just consists of translations and rotations, represented by
\begin{subequations}\label{AppenFormulae}
\be\label{piJi}
p_1,\, p_2,\, p_3,\, J_1= q_2 p_3-q_3 p_2,\, J_2= q_3 p_1-q_1 p_3,\, J_3= q_1 p_2-q_2 p_1.
\ee
An important conformal element is
\be\label{S-def}
S = q_1p_1+q_2p_2+q_3p_3,
\ee
satisfying $\{S,H^0\}=2 H^0$.

The Casimir of the rotation algebra is
\be\label{Jcas}
J^2 = J_1^2+J_2^2+J_3^2 = 2 (q_1^2+q_2^2+q_3^2) H^0-S^2.
\ee
The second representation, in terms of $H^0$ and $S$, is important when adding potentials, $H=H^0+U(q_1,q_2,q_3)$.

{\em Any} first integral of $H^0$ can be built from the ``Killing vectors'' (\ref{piJi}).  It is enough to consider {\em homogeneous} polynomials in the elements of (\ref{piJi}), representing higher order Killing tensors.  This is only true for flat and constant curvature metrics.  The second representation of $J^2$ in (\ref{Jcas}) is in terms of {\em conformal Killing vectors} and {\em tensors} and can be used in the context of conformally flat metrics (see \cite{f19-3,f19-2}).

Integrals for $H=H^0+U(q_1,q_2,q_3)$ have these ``Killing tensors'' as leading order parts.  A particularly important {\em quadratic} ``Killing tensor'' in this paper is
\be\label{X10}
X_1^0 = (p_1-p_2) J_3+(p_2-p_3) J_1+(p_3-p_1) J_2 = S (p_1+p_2+p_3) -2 (q_1+q_2+q_3) H^0.
\ee
Again, this second representation of $X_1^0$ is important when extending to $H=H^0+U(q_1,q_2,q_3)$ and also when considering conformally flat metrics.
\end{subequations}

\br[Coordinates $(Q_i,P_i)$]
Most of the calculations in this paper are performed within the $(Q_i,P_i)$ coordinates.  We use the same symbols $J_1=Q_2P_3-Q_3P_2$, etc for the rotations in these coordinates, but be warned that they are not invariant under the canonical transformation with (\ref{Qi}).  We also use $S$ to mean $\sum_iQ_iP_i$.  In this case the \underline{form} of $S$ \underline{is} invariant under this transformation, as shown in Section \ref{sec:QP-coords}.
\er

\subsection{Poisson Relations for Section \ref{sec:U2-Q-phi-F3}}\label{sec:appendix-pbs}

In Section \ref{sec:U2-Q-phi-F3} we considered the Poisson algebra associated with the Hamiltonian (\ref{H:U2-Q-phi-F3}).  We found that we could replace $X_1$, using $X_1=-X_5-X_6$, and that the 10 dimensional Poisson algebra consisted of 6 {\em quadratic} functions $H, X_2, \dots ,X_6$ and 4 cubic ones $X_7, \dots , X_{10}$.  The Poisson relations of the {\em quadratic} elements are simple, so given in Section \ref{sec:U2-Q-phi-F3}.  Here we consider the remaining 24 Poisson relations between quadratic and cubic functions, as well as the 6 Poisson relations between cubic ones.  This task is simplified by utilising a discrete symmetry of the Hamiltonian, which is invariant under the {\em involution} $\iota_{23}$ (which switches $Q_2$ and $Q_3$), if we extend this to act on 2 of the parameters: $k_2\leftrightarrow k_4$.  This enabled us to reflect this symmetry in our choice of functions, which satisfy:
$$
(H,X_2,X_3,X_4,X_5,X_6,X_7,X_8,X_9,X_{10},k_2,k_4)\mapsto(H,X_2,X_4,X_3,X_6,X_5,-X_7,-X_8,X_{10},X_9,k_4,k_2).
$$
As a result, if we have the formula for $\{X_3,X_9\}$, we can deduce the formula of $\{X_4,X_{10}\}$ by applying $\iota_{23}$. Introducing the notation $P_{ij}=\{X_i,X_j\}$, this means that $P_{39}$ and $P_{410}$ are paired.  On the other hand, since $P_{910}\mapsto -P_{910}$, we obtain nothing new.  This is depicted in the list below:
\bea
 && (P_{23},P_{24}),\quad (P_{25},P_{26}),\quad P_{27},\quad P_{28},\quad (P_{29},P_{210}),\quad P_{34},\quad (P_{35},P_{46}),\quad(P_{36},P_{45}),  \nn\\
 && (P_{37},P_{47}),\quad (P_{38},P_{48}),\quad (P_{39},P_{410}),\quad (P_{310},P_{49}),\quad P_{56}\quad (P_{57},P_{67}),\quad (P_{58},P_{68}),  \nn\\
 && (P_{59},P_{610}),\quad (P_{510},P_{69}),\quad P_{78},\quad (P_{79},P_{710}),\quad (P_{89},P_{810}),\quad P_{910} .  \nn
\eea
The additional, independent entries in the Poisson matrix are
\bea
&& P_{27}=2(X_4-X_3)X_2+2(k_2-k_4)(X_3+X_4),\quad P_{28}=4(X_6-X_5)X_2-4(k_2-k_4)(X_5+X_6),\nn\\
&& P_{29}=4(X_3X_6-X_4X_5), \;\;  P_{37}=-2X_3X_4+4\omega^2(X_2-k_2-k_4), \;\; P_{38}=4k_3(X_2-k_2-k_4)H-4X_4X_5,  \nn\\
&&  P_{39}=4k_3HX_3-8\omega^2X_5,\quad P_{310}=0, \quad  P_{57}=2k_3(X_2-k_2-k_4)H-2X_3X_6,\nn\\
&&   P_{58}=4k_1(k_2+k_4-X_2)H+3(X_2-k_2-k_4)X_3+2(X_2-2k_2-k_4)X_4-4X_5X_6,  \nn\\
&& P_{59}=-4(k_1X_3+k_3X_5)H+(3X_3+2X_4)X_3-4k_2\omega^2,\quad P_{510}=X_3X_4-2\omega^2(X_2-k_2-k_4),  \nn\\
&& P_{78}=4(X_6-X_5)X_7+2(X_3-X_4)X_8+4k_2X_{10}+4k_4X_9,\quad P_{79}=4k_3HX_7+2X_4X_9-4\omega^2X_8,  \nn\\
&& P_{89}=-4(2k_1X_7+k_3X_8)H+2(3X_3+2X_4)X_7+4X_6X_9,\quad P_{910}=4\omega^2X_7.     \nn
\eea
The ten integrals of the Poisson algebra are of rank 5 and satisfy the following five constraints
\bea
  && 2 k_3 H X_7 -2 \omega^2 X_8 -X_{10} X_3+X_4 X_9=0,   \nn\\
  && (2 k_1 H -X_3-X_4) X_7+k_3 H X_8 +X_{10} X_5-X_6 X_9=0,   \nn\\
  && 2(X_5-X_6) X_7+(X_4-X_3) X_8+(X_2-k_2-3 k_4) X_9+(X_2-3 k_2-k_4) X_{10}=0,   \nn\\
  && (X_2-k_2-k_4) (X_{10}-X_9)+(X_3+X_4)X_8 -2 (X_5+X_6)X_7 -2 k_4 X_9+2 k_2 X_{10} =0,   \nn\\
  && X_7^2-X_2 X_3 X_4+k_4 X_3^2+k_2 X_4^2+(k_2+k_4) X_3 X_4+\omega^2 \left((X_2-k_2-k_4)^2-4 k_2 k_4\right)=0.   \nn
\eea

\subsection{Poisson Relations for Section \ref{sec:ResOs}}\label{sec:appendix-pbs-flat}

The Hamiltonian (\ref{H:U2-Q-phi-F3}) reduces to its flat counterpart (\ref{HU2}), with condition (\ref{U2-F3B}), when $k_1=1,\, k_3=0$, along with the change of notation $(b_1,b_2)=\left(\frac{1}{2}k_2,\frac{1}{2}k_4\right)$.  Therefore, the Poisson algebra for the flat case is obtained from that of (\ref{H:U2-Q-phi-F3}), by this simple parametric reduction.

The definitions of $X_i$ and the Poisson relations, given in Section \ref{sec:U2-Q-phi-F3} are unchanged, but some of the $P_{ij}$, given above, are simplified.  Furthermore, the above five constraints can be replaced by the (mainly) simpler formulae:
\bea
  &&  X_4 X_9-X_{10} X_3 -2 \omega^2 X_8 =0, \quad (2 H -X_3-X_4) X_7 +X_{10} X_5-X_6 X_9=0,  \nn\\
  && X_7^2-X_2 X_3 X_4+k_4 X_3^2+k_2 X_4^2+(k_2+k_4) X_3 X_4+\omega^2 \left((X_2-k_2-k_4)^2-4 k_2 k_4\right)=0,   \nn\\
  &&  X_9^2+4\omega^2X_5^2+(X_3+X_4-2H)(X_3^2-4\omega^2 k_2)=0,  \nn\\
  &&  X_{10}^2+4\omega^2X_6^2+(X_3+X_4-2H)(X_4^2-4\omega^2 k_4)=0.  \nn
\eea


\end{document}